\documentclass[
prc,
showpacs,amssymb,superscriptaddress,aps
]{revtex4-2}
\usepackage{graphicx}
\usepackage{color}
\input{epsf}
\usepackage{amsmath,amssymb}
\usepackage{bm}
\usepackage{times}
\usepackage{dcolumn}
\newcommand{\dalm}{\kern1pt\vbox{\hrule height 0.9pt\hbox{\vrule width 0.9pt
\hskip 2.5pt\vbox{\vskip 5.5pt}\hskip 3pt\vrule width 0.3pt}\hrule height 0.3pt}
\kern1pt}

\newcommand{\lsim}{\, \, \raisebox{-0.8ex}{$\stackrel{\textstyle <}{\sim}$ }}
 
\preprint{RIKEN-iTHEMS-Report-23}
\begin{document}
%
%
%
\title{Empirical neutron star mass formula based on experimental observables}
\author{Hajime Sotani}
\email{sotani@yukawa.kyoto-u.ac.jp}
\affiliation{Astrophysical Big Bang Laboratory, RIKEN, Saitama 351-0198, Japan}
\affiliation{Interdisciplinary Theoretical \& Mathematical Science Program (iTHEMS), RIKEN, Saitama 351-0198, Japan}
\author{Tomoya Naito}
\affiliation{Interdisciplinary Theoretical \& Mathematical Science Program (iTHEMS), RIKEN, Saitama 351-0198, Japan}
\affiliation{Department of Physics, Graduate School of Science, The University of Tokyo, Tokyo 113-0033, Japan}
\date{\today}
%
%
\begin{abstract}
  We derive the empirical formulae expressing the mass and gravitational redshift of a neutron star,
  whose central density is less than threefold the nuclear saturation density,
  as a function of the neutron-skin thickness or the dipole polarizability of $ {}^{208} \mathrm{Pb} $ or $ {}^{132} \mathrm{Sn} $,
  especially focusing on the 8
  Skyrme-type effective interactions.
  The neutron star mass and its gravitational redshift can be estimated within $ \approx 10 \, \% $ errors with our formulae,
  while the neutron star radius is also expected within a few $\%$ errors by combining the derived formulae.
  Owing to the resultant empirical formulae, we find that the neutron star mass and radius are more sensitive to the neutron-skin thickness of $ {}^{208} \mathrm{Pb} $ than the dipole polarizability of $ {}^{208} \mathrm{Pb} $ or $ {}^{132} \mathrm{Sn} $. 
\end{abstract}
%
\maketitle
%
%
\section{Introduction}
\label{sec:I}
\par
The death of a massive star would be the brightest moment in its life.
This is known as a supernova explosion.
Through such an explosion, a neutron star would come into the world as a massive remnant.
The neutron star realizes extreme states, which are almost impossible to reproduce on Earth \cite{ST83}.
The density inside the star easily exceeds the standard nuclear density, $\rho_0$, and may reach several times larger, if not more, than $\rho_0$, depending on the equation of state (EOS) for neutron star matter.
The magnetic and gravitational fields inside/around the star also become much stronger than those observed in our solar system.
So, the observations of neutron stars and/or their phenomena might inversely tell us the aspect of such extreme states.
In particular, the constraints from the observations on the neutron star mass and radius are directly associated with the validity of the EOSs.
For instance, the discoveries of massive neutron stars could exclude some of soft EOSs, whose expected maximum mass is less than the observed mass \cite{D10,A13,C20,F21}.
The gravitational wave observations from the binary neutron star mergers, GW170817 and GW190425, enable us to estimate the mass and radius of neutron stats before mergers \cite{gw170817,gw190425}.
The observation of GW170817 also gives us information on the dimensionless tidal deformability, which leads to the constraint on the radius of $1.4M_{\odot}$ neutron star, i.e., $R_{1.4}\lesssim 13.6 \, \mathrm{km}$ \cite{Annala18}.
Furthermore, owing to the relativistic effect, i.e., the light bending due to the strong gravitational field induced by the neutron star, one could primarily constrain the neutron star compactness (the ratio of the mass to radius) by carefully observing the pulsar light curves (e.g., \cite{PFC83,LL95,PG03,PO14,SM18_1,SM18_2,Sotani20a}).
In fact, the x-ray observations with the Neutron star Interior Composition ExploreR (NICER) succeed to put the constraint on the neutron star mass and radius, i.e., PSR J0030+0451 \cite{Riley19,Miller19} and PSR J0740+6620 \cite{Riley21,Miller21}.
In general, astronomical observations tend to constrain the properties of neutron stars in a higher-density region. 
\par
On the other hand, via nuclear experiments performed on Earth,
one could achieve information on the relatively lower-density region.
Since any EOS model can be characterized with its own nuclear EOS parameters,
the constraint on such parameters through the terrestrial experiments would partially restrict on the EOS for neutron star matter,
which enables us to narrow the allowed region of neutron star mass and radius.
For example, the observation of the ratio of the positive charged pions to the negative ones in the decay process of $\Delta$ isobars into nucleons with pions,
using the isotope beams provided by the Radioactive Isotope Beam Factory (RIBF) at RIKEN in Japan,
constrains the density-dependence of symmetry energy, $L$, i.e., $ 42 \le L \le117 \, \mathrm{MeV} $ (S$\pi$RIT, e.g., \cite{SpiRIT}).
The estimation of the neutron-skin thickness of $ {}^{208} \mathrm{Pb} $,
using the parity-violating asymmetry of the elastic electron scattering cross section measured at the Thomas Jefferson National Accelerator Facility in Virginia,
also constrains $L$ as $ L = 106 \pm 37 \, \mathrm{MeV} $ (PREX-II \cite{PREXII}).
We note that these two constraints on $L$ seem to be relatively large values compared to the fiducial value of $L$, i.e., $ L \simeq 60 \pm 20 \, \mathrm{MeV} $ \cite{Vinas14,BALi19}.
Anyway, by using these constraints on the saturation parameters, one can discuss the expected neutron star mass and radius by using the empirical formulae expressing the neutron star mass and its gravitational redshift as a function of the suitable combination of nuclear saturation parameters \cite{SIOO14,ST22} or those for asymmetric nuclear matter \cite{SO22}.
In such a way, the astronomical observations and experimental constraints must be complementary approaches for fixing the EOS of neutron star matter \cite{SNN22}.
\par
Since the nuclear saturation parameters cannot be directly measured, one has to evaluate them using the experimental data strongly associated with the saturation parameters.
Up to now, several strong correlations between the experimental observables and saturation parameters have been found theoretically, which helps us to estimate the saturation parameters from the experiments.
However, one might have to say that these correlations are still incomplete.
Due to the theoretical uncertainties in the correlations, it is not always true that the constraint on the saturation parameters becomes more severe, even though the accuracy in experiments would be improved well~\cite{
  Erler2013Phys.Rev.C87_044320,
  Kurasawa2022Prog.Theor.Exp.Phys.2022_023D03,
  Reinhard2022Phys.Rev.C105_L021301,
  Reinhard2022Phys.Rev.Lett.129_232501,
  Naito2022Phys.Rev.C106_L061306,  
  NCLRS23}.
For instance,
the estimation of the $ L $ parameter using the parity asymmetry of the polarized electron scattering cross section of $ {}^{208} \mathrm{Pb} $ is strongly model dependent; 
even based on the same experimental data,
different groups estimate different values of the $ L $ parameter~\cite{PREXII,PREXII-Reanalysis}.
So, in this study, we will consider to derive the empirical formulae expressing the neutron star mass and its gravitational redshift directly using the experimental observables, instead of the nuclear saturation parameters as in Refs.~\cite{SIOO14,ST22,SO22}.
For this purpose, we adopt the 8 EOS models with the Skyrme energy density functionals, where 5 of them are SLy models (see the next section for details).
  So, our discussion can be done in a relatively small parameter range, i.e., $40\lsim L\lsim 80 \, \mathrm{MeV} $.
  In addition, the empirical relations will be derived by using the neutron star models, whose central density is less than threefold the saturation density.
  Since the corresponding neutron star masses are at most $\approx 1M_{\odot}$,
  one may not be able to directly discuss the real observations of neutron star mass with our empirical relations.
\par
This manuscript is organized as follows.
In Sec.~\ref{sec:EOS}, we briefly mention the EOSs considered in this study and the experimental observables estimated from each EOS model.
In Sec.~\ref{sec:NS}, we derive the empirical formulas for the neutron star mass and its gravitational redshift, where we also show the relative accuracy in the estimations from our empirical formulae.
Then, in Sec.~\ref{sec:MR}, we discuss the neutron star mass and radius expected from the resultant empirical formulae, together with the constraints from the astronomical and experimental observations.
Finally, we conclude this study in Sec.~\ref{sec:Conclusion}.
Unless otherwise mentioned, we adopt geometric units in the following, $c=G=1$, where $c$ and $G$ denote the speed of light and the gravitational constant, respectively.
%
%
\begin{table*}[tb]
  \centering
  \caption{
    EOS parameters adopted in this study,
    $n_0$, $E/A$, $K_0$, $S_0$, $L$, $K_{\mathrm{sym}}$,
    are listed,
    while $\eta$ is a specific combination with them given by
    $\eta = \left(K_0 L^2\right)^{1/3}$.
    In addition, the maximum mass expected with each EOS is also listed~\cite{
      Compose}.}
  \label{tab:EOS2}
  \begin{ruledtabular}
    \begin{tabular}{lddddddddl}
      EOS     & \multicolumn{1}{c}{$n_0$ ($ \mathrm{fm}^{-3} $)} & \multicolumn{1}{c}{$w_0$ ($ \mathrm{MeV} $)} & \multicolumn{1}{c}{$K_0$ ($ \mathrm{MeV} $)} & \multicolumn{1}{c}{$S_0$ ($ \mathrm{MeV} $)} & \multicolumn{1}{c}{$L$ ($ \mathrm{MeV} $)} & \multicolumn{1}{c}{$K_{\mathrm{sym}}$ ($ \mathrm{MeV} $)} & \multicolumn{1}{c}{$\eta$ ($ \mathrm{MeV} $)} & \multicolumn{1}{c}{$M_{\mathrm{max}} $ ($ M_{\odot} $)} & \multicolumn{1}{c}{Ref.} \\
      \hline
      LNS5    &  0.15992  & -15.57 & 240.2 & 29.15 &  50.94 &-119.1  & 85.43 & 1.97 & \cite{Gambacurta2011Phys.Rev.C84_024301} \\   
      SKa     &  0.15535  & -15.99 & 263.1 & 32.91 & 74.62 & -78.45  & 113.6 & 2.22 & \cite{Koehler1976Nucl.Phys.A258_301} \\   
      SkMp    &  0.15704  & -15.56 & 230.9 & 29.89 & 70.31 & -49.82  & 104.5 & 2.11 & \cite{Bennour1989Phys.Rev.C40_2834} \\  
      SLy2    &  0.16053  & -15.99 & 229.9 & 32.00 & 47.46 & -115.1  & 80.30 & 2.06 & \cite{Chabanat1995UniversiteClaudeBernardLyon1_PhD} \\  
      SLy4    &  0.15954  & -15.97 & 229.9 & 32.00 & 45.96 & -119.7  & 78.60 & 2.06 & \cite{Chabanat1998Nucl.Phys.A635_231} \\   
      SLy5    &  0.16034  & -15.98 & 229.9 & 32.03 & 48.27 & -112.3  & 81.22 & 2.02 & \cite{Chabanat1998Nucl.Phys.A635_231} \\   
      SLy9    &  0.15117  & -15.79 & 229.8 & 31.98 & 54.86 & -81.42  & 88.43 & 2.16 & \cite{Chabanat1995UniversiteClaudeBernardLyon1_PhD} \\   
      SLy230a &  0.15997  & -15.99 & 229.9 & 31.99 & 44.32 & -98.22  & 76.72 & 2.11 & \cite{Chabanat1997Nucl.Phys.A627_710} \\
    \end{tabular}
  \end{ruledtabular}
\end{table*}
%
\section{EOS for neutron star matter and experimental variables}
\label{sec:EOS}
\par
In order to construct a cold neutron star model, one has to prepare the EOS for neutron star matter with zero temperature, which satisfies the charge neutral and beta-equilibrium conditions.
Up to now, many EOS models have been proposed, but the EOSs
whose thermodynamical properties are opened as a tabulated format (or analytic expression)
are not so much.
In this study, we especially focus on the 8 EOS models constructed with the Skyrme-type effective interaction~\cite{
  Vautherin1972Phys.Rev.C5_626,
  Bender2003Rev.Mod.Phys.75_121},
which are commonly accepted and give us a reasonable nuclear structure.
\par
These Skyrme interactions contain $ 10 $ (or $ 11 $) parameters,
which are determined to reproduce experimental data of several nuclear properties,
such as binding energies and charge radii,
and/or nuclear matter properties.
Different model, i.e., different Skyrme interaction, adopts different criteria to determine the parameters,
which lead to different calculation results especially for open-shell or exotic nuclei~\cite{
  Danielewicz2009Nucl.Phys.A818_36,
  Yoshida2013Prog.Theor.Exp.Phys.2013_113D02,
  Danielewicz2014Nucl.Phys.A922_1,
  Danielewicz2017Nucl.Phys.A958_147,
  Naito:2022vnz}.
So far, there are no standard or ultimate criteria to optimize the parameters of the Skyrme interaction.
Here, we adopt only the EOS models taking into account the one-body center-of-mass correction without the two-body one~\cite{
  Bender2000Eur.Phys.J.A7_467}.
In addition, we select only the EOS models, with which the expected maximum mass exceeds (or is comparable to) the $2M_{\odot}$ neutron star observations.
The EOS data are taken from the public source in CompStar Online Supernovae Equations of State (CompoOSE \cite{Compose}).
The EOS models adopted in this study are listed in Table \ref{tab:EOS2}.
Once the EOS is prepared, the neutron star model is determined by integrating the Tolman-Oppenheimer-Volkoff (TOV) equations. 
\par
For any EOS model, one can express the bulk energy per nucleon for the zero-temperature uniform nuclear matter as a function of the baryon number density,
$ n_{\mathrm{b}} = n_n + n_p $,
and an asymmetric parameter defined as
$ \alpha = \left( n_n - n_p \right)/n_{\mathrm{b}} $ with neutron (proton) number density,
$ n_n $ ($ n_p $),
in the vicinity of saturation density, $ n_0 \simeq 0.16 \, \mathrm{fm}^{-3} $,
for a symmetric nuclear matter as
\begin{widetext}
  \begin{equation}
    \label{eq:enegy}
    \frac{E}{A} \left( n_{\mathrm{b}}, \alpha \right)
    =
    w_0
    +
    \frac{K_0}{2}
    u^2
    +
    \mathcal{O} \left( u^3 \right) 
    +
    \left[
      S_0
      +
      Lu
      +
      \frac{K_{\mathrm{sym}}}{2}
      u^2
      +
      \mathcal{O} \left( u^3 \right)
    \right]
    \alpha^2 
    +
    \mathcal{O} \left( \alpha^3 \right), 
  \end{equation}
\end{widetext}
where
$ u = \left( n_{\mathrm{b}} - n_0 \right) / \left( 3n_0 \right) $
and the coefficient of $ \alpha^2 $ corresponds to the nuclear symmetry energy.
The saturation parameters appearing in Eq.~(\ref{eq:enegy}) for the EOS models adopted in this study are listed in Table \ref{tab:EOS2}.
As mentioned below, in this study we focus on $ S_0 $, which has been constrained well from the experiments, i.e., $ S_0 \approx 31.6 \pm 2.7 \, \mathrm{MeV} $ \cite{BALi19}.
\par
On the other hand, using each model, i.e., Skyrme interaction, one can estimate the experimental observables,
such as the ground-state energy, $ E_{\mathrm{gs}} $,
the neutron-skin thickness, $ \Delta R_n $,
the dipole polarizability, $ \alpha_{\mathrm{D}} $,
and the energy of isoscalar giant monopole resonance (ISGMR), $ E_{\mathrm{ISGMR}} $,
for specific atomic nuclei.
It is known that $ \Delta R_n $ and $ \alpha_{\mathrm{D}} $ are correlated with the $ L $ parameter~\cite{
  RocaMaza11,
  RocaMaza13},
while $ E_{\mathrm{ISGMR}} $ is correlated with the $ K_0 $ parameter~\cite{
  Blaizot1980Phys.Rep.64_171,
  Piekarewicz2007Phys.Rev.C76_031301,
  Garg2018Prog.Part.Nucl.Phys.101_55,
  Li:2022suc}.
They are calculated by using an open-source code for the spherical Hartree-Fock and the random phase approximation (RPA) calculation named \textsc{skyrme\_rpa}~\cite{
  Colo2013Comput.Phys.Commun.184_142}.
Since $ {}^{208} \mathrm{Pb} $ and $ {}^{132} \mathrm{Sn} $ are doubly magic, 
we can safely calculate by assuming the spherical symmetry without the pairing correlation.
Owing to the nature of the spherical symmetry,
only the radial wave function is calculated,
where we consider the box size of $ 0 < r < 20 \, \mathrm{fm} $  with a $ 0.1 \, \mathrm{fm} $ mesh.
Starting from the Hartree-Fock ground state,
the RPA calculation is performed,
where the cut-off energy for unoccupied single-particle orbitals
is $ 60 \, \mathrm{MeV} $ for $ {}^{132} \mathrm{Sn} $
and $ 80 \, \mathrm{MeV} $ for $ {}^{208} \mathrm{Pb} $.
The obtained strength functions are smeared with the Lorentzian function with a $ 1.0 \, \mathrm{MeV} $ width.
  We note that $\Delta R_n$ and $\alpha_{\text{D}}$ for
  $ {}^{48} \mathrm{Ca} $ and $ {}^{208} \mathrm{Pb} $ have been measured as
  $ \Delta R_n^{\text{Ca}} = 0.168_{-0.028}^{+0.025} \, \mathrm{fm} $~\cite{Zenihiro:2018rmz},
  $ \Delta R_n^{\text{Ca}} = 0.121 \pm 0.026 \text{(exp)} \pm 0.024 \text{(model)} \, \mathrm{fm} $~\cite{PhysRevLett.129.042501};
  $ \alpha_{\text{D}}^{\text{Ca}} = 2.07(22) \, \mathrm{fm}^3 $~\cite{PhysRevLett.118.252501};
  $\Delta R_n^{\text{Pb}} = 0.211_{-0.063}^{+0.054} \, \mathrm{fm} $~\cite{Zenihiro2010Phys.Rev.C82_044611}, 
  $\Delta R_n^{\text{Pb}} = 0.283 \pm 0.071 \, \mathrm{fm} $~\cite{PREXII}; and
  $\alpha_{\text{D}}^{\text{Pb}} = 20.1 (6) \, \mathrm{fm}^3 $~\cite{
    Tamii2011Phys.Rev.Lett.107_062502,
    10.1143/PTPS.196.166},
  while $\Delta R_n$ and $\alpha_{\text{D}}$ for $ {}^{132} \mathrm{Sn} $ are not known in experiments.
  However, it is discussed in Refs.~\cite{
    Perera2021Phys.Rev.C104_064313,
    Sagawa2022Phys.Lett.B829_137072,
    Yang:2022tjf,
    Minato:2022kes,
    Naito:2022vnz,
    NCLRS23}
  that beyond-mean-field effects may be indispensable to calculate properties of $ {}^{40} \mathrm{Ca} $ and $ {}^{48} \mathrm{Ca} $ consistently,
  while the mean-field calculation is used in this paper.
  Hence, in this study,
  we use properties of $ {}^{132} \mathrm{Sn} $ and $ {}^{208} \mathrm{Pb} $
  obtained by the RPA calculation.
\par
The Hartree-Fock calculation provides the ground-state energy and density.
Using the ground-state proton and neutron densities,
the neutron-skin thickness can be calculated by
$ \Delta R_n = R_n - R_p $,
where $ R_p $ ($ R_n $) is the proton (neutron) root-mean-square radius.
The RPA calculation provides the strength function.
The energy of ISGMR, $ E_{\mathrm{ISGMR}} $, corresponds to the peak of the strength function of the isoscalar monopole resonance.
The dipole polarizability is related to the isovector dipole resonance,
which can be calculated by~\cite{
  RocaMaza13}
\begin{equation}
  \alpha_{\mathrm{D}}
  =
  \frac{8 \pi e^2}{9}
  m_{-1} \left( E1 \right),
\end{equation}
where $ m_{-1} \left( E1 \right) $ the inverse energy weighted sum rule of the isovector dipole resonance~\cite{
  Zhang2018Phys.Lett.B777_73}.
These results for $ {}^{208} \mathrm{Pb} $ and $ {}^{132} \mathrm{Sn} $ calculated with the Skyrme interactions considered in this study are listed in Tables \ref{tab:208Pb} and \ref{tab:132Sn}.
%
\begin{table}[tb]
  \centering
  \caption{
    Estimation of ground-state energy, $ E_{\mathrm{gs}} $,
    neutron-skin thickness, $ \Delta R_n $,
    dipole polarizability, $ \alpha_{\mathrm{D}} $,
    and energy of ISGMR, $ E_{\mathrm{ISGMR}} $,
    for $ {}^{208} \mathrm{Pb} $, using various Skyrme interactions.} 
  \label{tab:208Pb}
  \begin{ruledtabular}
    \begin{tabular}{ldddd}
      EOS & \multicolumn{1}{c}{$ E_{\mathrm{gs}}^{\mathrm{Pb}} $ ($ \mathrm{MeV} $)} & \multicolumn{1}{c}{$ \Delta R_n^{\mathrm{Pb}} $ ($ \mathrm{fm} $)} & \multicolumn{1}{c}{$ \alpha_{\mathrm{D}}^{\mathrm{Pb}} $ ($ \mathrm{fm}^3 $)} & \multicolumn{1}{c}{$ E_{\mathrm{ISGMR}}^{\mathrm{Pb}} $ ($ \mathrm{MeV} $)} \\
      \hline
      LNS5    & -1625.6 & 0.1577 & 21.47 & 13.97 \\   
      SKa     & -1636.5 & 0.2114 & 22.36 & 14.09 \\   
      SkMp    & -1636.9 & 0.1958 & 23.90 & 13.71 \\  
      SLy2    & -1635.9 & 0.1637 & 19.62 & 13.51 \\  
      SLy4    & -1636.0 & 0.1597 & 19.81 & 13.57 \\   
      SLy5    & -1636.4 & 0.1624 & 19.92 & 13.61 \\   
      SLy9    & -1630.3 & 0.1716 & 20.93 & 13.28 \\   
      SLy230a & -1635.9 & 0.1525 & 19.20 & 13.55 \\   
    \end{tabular}
  \end{ruledtabular}
\end{table}
%
\begin{table}[tb]
  \centering
  \caption{
    Same as in Table \ref{tab:208Pb}, but for $ {}^{132} \mathrm{Sn} $.}
  \label{tab:132Sn}
  \begin{ruledtabular}
    \begin{tabular}{ldddd}
      EOS & \multicolumn{1}{c}{$ E_{\mathrm{gs}}^{\mathrm{Sn}} $ ($ \mathrm{MeV} $)} & \multicolumn{1}{c}{$ \Delta R_n^{\mathrm{Sn}} $ ($ \mathrm{fm} $)} & \multicolumn{1}{c}{$ \alpha_{\mathrm{D}}^{\mathrm{Sn}} $ ($ \mathrm{fm}^3 $)} & \multicolumn{1}{c}{$ E_{\mathrm{ISGMR}}^{\mathrm{Sn}} $ ($ \mathrm{MeV} $)} \\
      \hline
      LNS5    & -1108.4 & 0.2162 & 10.62 & 16.25 \\   
      SKa     & -1105.8 & 0.2736 & 11.22 & 16.30 \\   
      SkMp    & -1119.7 & 0.2589 & 11.86 & 16.04 \\  
      SLy2    & -1103.4 & 0.2252 & 9.796 & 15.57 \\  
      SLy4    & -1103.7 & 0.2214 & 9.906 & 15.63 \\   
      SLy5    & -1104.1 & 0.2245 & 9.964 & 15.63 \\   
      SLy9    & -1097.5 & 0.2345 & 10.49 & 15.74 \\   
      SLy230a & -1103.1 & 0.2132 & 9.568 & 15.66 \\   
    \end{tabular}
  \end{ruledtabular}
\end{table}
%
\begin{figure}[tb]
  \centering
  \includegraphics[width=1.0\linewidth]{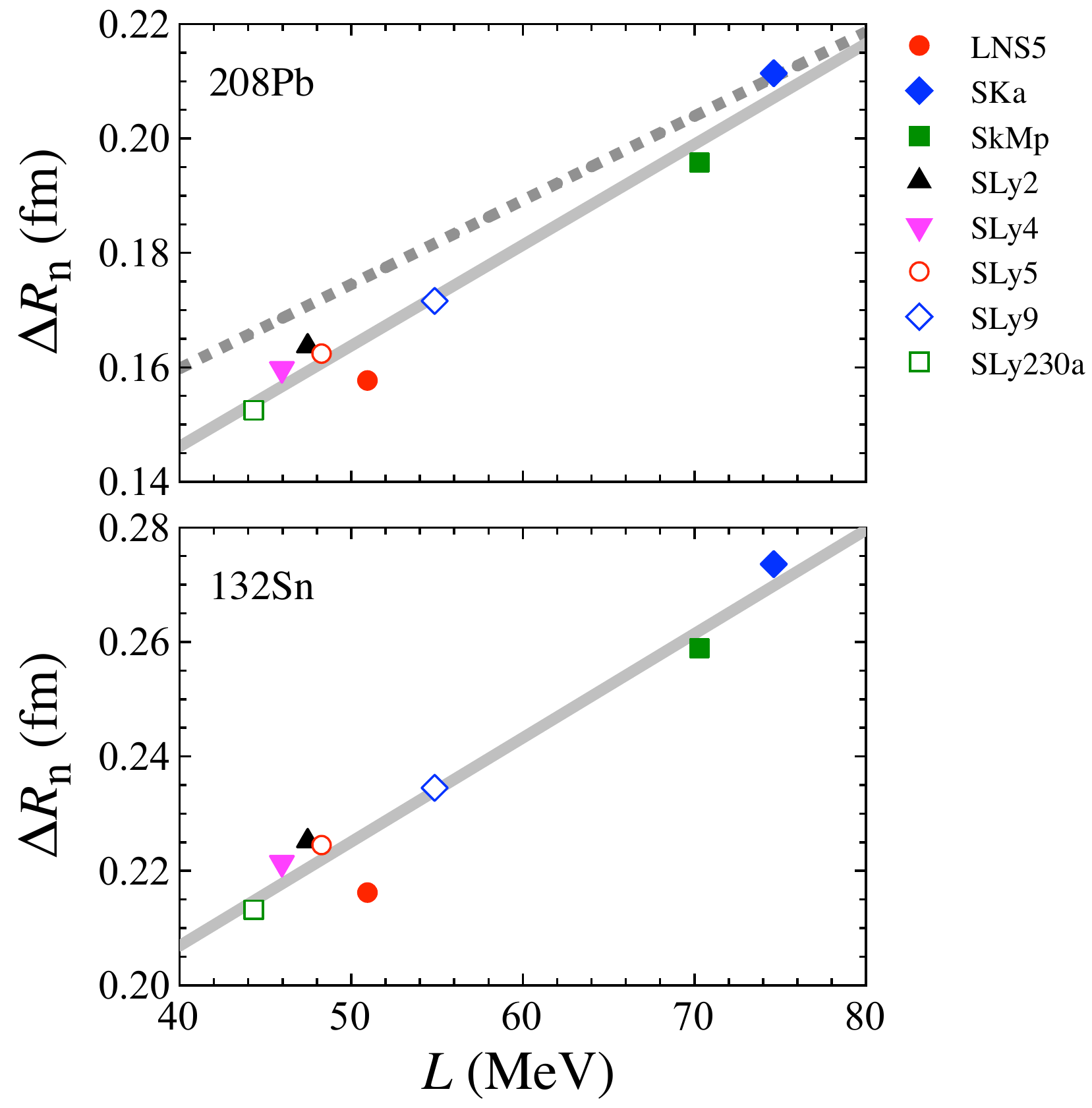}
  \caption{
    Neutron-skin thickness, $\Delta R_n$, expected with each EOS model is shown as a function of the corresponding value of $ L $.
    The top and bottom panels correspond to the result for $ {}^{208} \mathrm{Pb} $ and $ {}^{132} \mathrm{Sn} $.
    The fitting lines given by Eqs.~(\ref{eq:fitting_DRnPb1}) and  ~(\ref{eq:fitting_DRnSn}) are shown with the solid lines, while the fitting line derived in Ref.~\cite{RocaMaza11} is also shown with the dotted line in the top panel.}
  \label{fig:DRn-L}
\end{figure}
%
\begin{figure}[tb]
  \centering
  \includegraphics[width=1.0\linewidth]{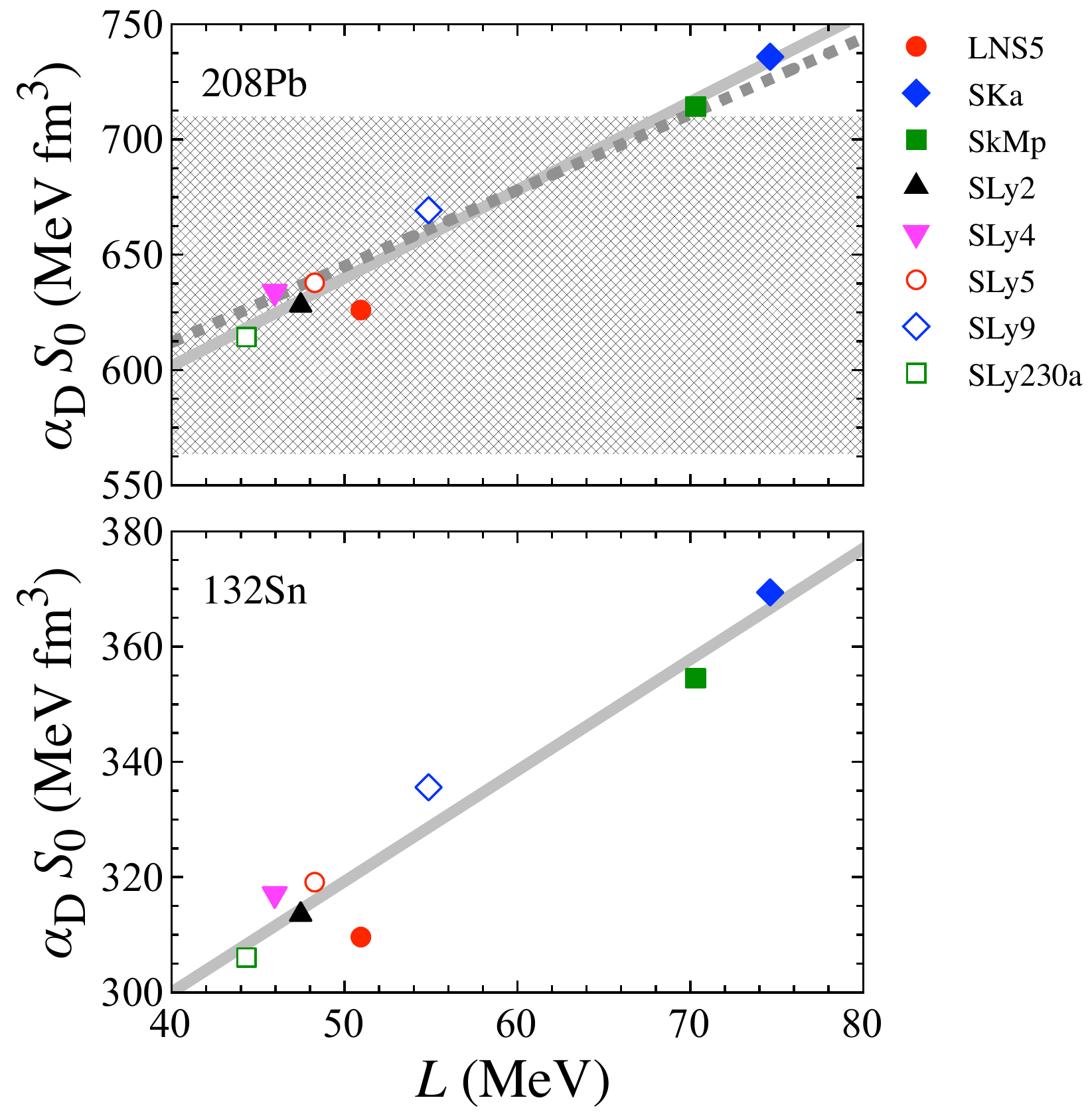}
  \caption{
    Dipole polarizability multiplied with $ S_0 $ for each EOS model is shown as a function of $ L $,
    where the top and bottom panels correspond to the result for $ {}^{208} \mathrm{Pb} $ and $ {}^{132} \mathrm{Sn} $.
    The fitting lines given by Eqs.~(\ref{eq:fitting_alphaS0Pb1}) and  ~(\ref{eq:fitting_alphaS0Sn}) are shown with the solid lines, while the fitting line derived in Ref.~\cite{RocaMaza13} is also shown with the dotted line in the top panel.
    For reference, the experimental value of $\alpha_{\mathrm{D}}^{\mathrm{Pb}}
    S_0$ is shown with the shaded region in the top panel, using the experimental value  $\alpha_{\text{D}}^{\text{Pb}} = 20.1(6) \, \mathrm{fm}^3 $~\cite{10.1143/PTPS.196.166} together with  $ S_0 \approx 31.6 \pm 2.7 \, \mathrm{MeV} $~\cite{BALi19} as the fiducial value of $S_0$.}
  \label{fig:alphaS-L}
\end{figure}
%
\par
Up to now, it has been already shown that the neutron-skin thickness and dipole polarizability multiplied with $S_0$ for $ {}^{208} \mathrm{Pb} $ are respectively associated with $L$ \cite{RocaMaza11,RocaMaza13},
such as
\begin{subequations}
  \begin{align}
    \Delta R_n^{\mathrm{Pb}}
    \, \left( \mathrm{fm} \right)
    & =
      0.101 + 0.147 L_{100},
      \label{eq:fitting_DRnPb} \\
    \alpha_{\mathrm{D}}^{\mathrm{Pb}}
    S_0
    \, \left( \mathrm{MeV} \, \mathrm{fm}^3 \right)
    & =
      480 + 330L_{100},
      \label{eq:fitting_alphaS0Pb}
  \end{align}
\end{subequations}
where $ L_{100} = L / \left( 100 \, \mathrm{MeV} \right) $.
In a similar way,
we can confirm the same properties of $ {}^{208} \mathrm{Pb} $ and $ {}^{132} \mathrm{Sn} $ estimated with EOS models adopted in this study are also strongly associated with $L$, i.e.,
\begin{subequations}
  \begin{align}
    \Delta R_n^{\mathrm{Pb}}
    \, \left( \mathrm{fm} \right)
    & =
      0.0757 + 0.176 L_{100},
      \label{eq:fitting_DRnPb1} \\
    \Delta R_n^{\mathrm{Sn}}
    \, \left( \mathrm{fm} \right)
    & =
      0.134 + 0.182L_{100},
      \label{eq:fitting_DRnSn} \\
    \alpha_{\mathrm{D}}^{\mathrm{Pb}}
    S_0
    \, \left( \mathrm{MeV} \, \mathrm{fm}^3 \right)
    & =
      449 + 382 L_{100},
      \label{eq:fitting_alphaS0Pb1} \\
    \alpha_{\mathrm{D}}^{\mathrm{Sn}}
    S_0
    \, \left( \mathrm{MeV} \, \mathrm{fm}^3 \right)
    & =
      232 + 177L_{100}.
      \label{eq:fitting_alphaS0Sn}
  \end{align}
\end{subequations}
These fitting lines are shown in Figs.~\ref{fig:DRn-L} and \ref{fig:alphaS-L} together with the concrete values estimated with each EOS model.
From these figures, we observe that the neutron-skin thickness of $ {}^{208} \mathrm{Pb} $ relatively deviates from the empirical relation given by Eq.~(\ref{eq:fitting_DRnPb}),
while the values of $ \alpha_{\mathrm{D}} S_0 $ of $ {}^{208} \mathrm{Pb} $ are more or less consistent with the prediction with Eq.~(\ref{eq:fitting_alphaS0Pb}),
using the EOS models adopted in this study.
In the top panel of Fig.~\ref{fig:alphaS-L}, we also show the experimental value of $\alpha_{\mathrm{D}}^{\mathrm{Pb}}
S_0$ with the shaded region, using the experimental value  $\alpha_{\text{D}}^{\text{Pb}} = 20.1(6) \, \mathrm{fm}^3 $~\cite{10.1143/PTPS.196.166} together with  $ S_0 \approx 31.6 \pm 2.7 \, \mathrm{MeV} $~\cite{BALi19} as the fiducial value of $S_0$. 
%
\section{Neutron star mass formula}
\label{sec:NS}
\par
We have already shown that the mass, $M$, and gravitational redshift, $z$,
of a low-mass neutron star can be expressed well as a function of the normalized central density,
$ u_{\mathrm{c}} = \rho_{\mathrm{c}} / \rho_0 $ using the central density $ \rho_{\mathrm{c}} $,
and the suitable combination of the saturation parameters \cite{SIOO14,ST22,SO22}.
Here, a low-mass neutron star means that the central density is less than a few times the nuclear saturation density.
Since the gravitational redshift is expressed with $M$ and the stellar radius, $R$,
as $ z = \left( 1 - 2 M/R \right)^{-1/2} - 1 $,
one can estimate the neutron star mass and radius from these empirical formulae for $M$ and $z$,
once the saturation parameters would be constrained.
In this study, we consider similar possibilities for expressing $M$ and $z$ directly with the experimental observables,
such as the neutron-skin thickness and the dipole polarizability, instead of the saturation parameters~\footnote{
  In this study, we focus only on the neutron-skin thickness and the dipole polarizability as the experimental observables.
  But, it has been shown that the parity-violating asymmetry for $ {}^{208} \mathrm{Pb} $ is also strongly associated with the neutron-skin thickness almost independently of the EOS models \cite{RocaMaza11}.
  So, one may derive the empirical formula expressing the neutron star mass and gravitational redshift, using the parity-violating asymmetry.}.
%
\subsection{Empirical relations with $ \Delta R_n $}
\label{sec:NSa}
\par
\begin{figure*}[tb]
  \centering
  \includegraphics[width=1.0\linewidth]{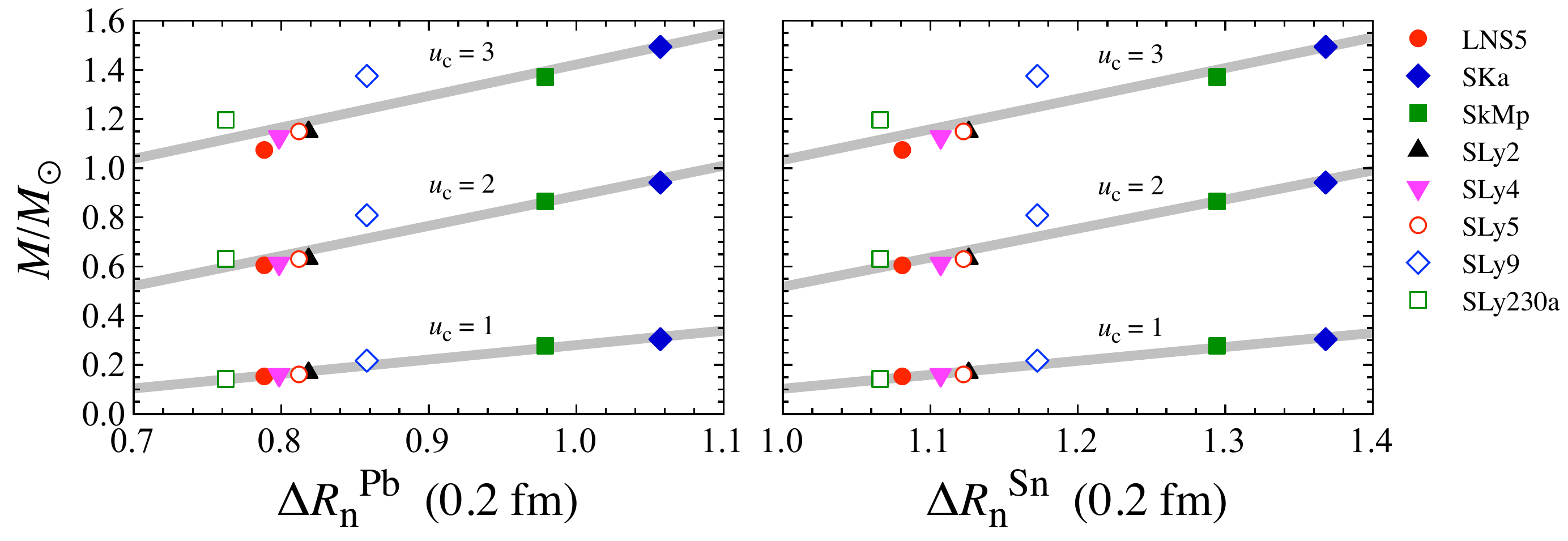}
  \caption{
    The mass of the neutron stars constructed with each EOS model is shown as a function of
    $ \Delta R_n^{\mathrm{Pb}} $
    in the left panel and
    $ \Delta R_n^{\mathrm{Sn}} $
    in the right panel,
    where the neutron star mass for $ u_{\mathrm{c}} = 1 $, $ 2 $, and $ 3 $ are shown.
    The fitting lines are given by Eqs.~(\ref{eq:fit_m_DR_Pb}) and~(\ref{eq:fit_m_DR_Sn}).}
  \label{fig:M-DR}
\end{figure*}
\begin{figure}[bp]
  \centering
  \includegraphics[width=1.0\linewidth]{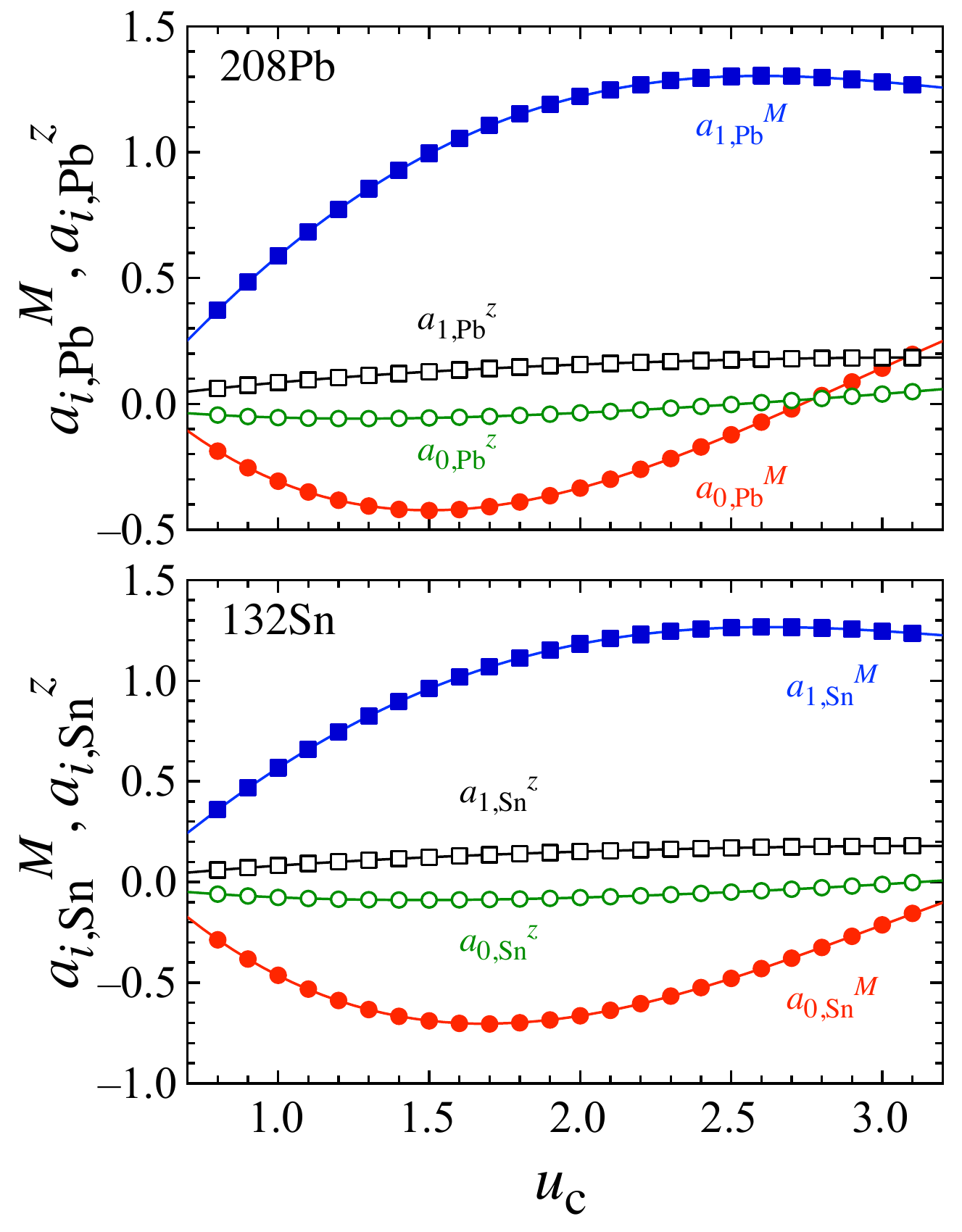}
  \caption{
    The coefficients in
    Eqs.~(\ref{eq:fit_m_DR_Pb}), (\ref{eq:fit_m_DR_Sn}),
    (\ref{eq:fit_z_DR_Pb}), and (\ref{eq:fit_z_DR_Sn})
    are shown as a function of $ u_{\mathrm{c}} $,
    where the top and bottom panels correspond to the coefficients in the formulae with the data of
    $ {}^{208} \mathrm{Pb} $ and $ {}^{132} \mathrm{Sn} $, respectively,
    while the solid lines denote the fitting lines given by
    Eqs.~(\ref{eq:M-DR-Pb-a}), (\ref{eq:M-DR-Sn-a}),
    (\ref{eq:z-DR-Pb-a}), and (\ref{eq:z-DR-Sn-a}).}
  \label{fig:a_dRM}
\end{figure}

First, we consider the possibility of deriving the empirical formulae for $M$ and $z$ with the neutron-skin thickness, $\Delta R_n^{\mathrm{Pb}}$ and  $\Delta R_n^{\mathrm{Sn}}$.
We eventually find a correlation between the mass of the neutron star constructed with a fixed central density and the neutron-skin thickness weakly depending on the EOS models.
In Fig.~\ref{fig:M-DR}, we show this feature, where the left and right panels correspond to the results with $ {}^{208} \mathrm{Pb} $ and $ {}^{132} \mathrm{Sn} $
for the neutron star models with $ u_{\mathrm{c}} = 1 $, $ 2 $, and $ 3 $.
In this figure, the solid lines denote the resultant linear fitting, given by
\begin{subequations}
  \label{eq:fit_m_DR}
  \begin{align}
    \frac{M}{M_{\odot}}
    & =
      a_{0, {\mathrm{Pb}}}^M
      +
      a_{1, {\mathrm{Pb}}}^M
      \left( \frac{\Delta R_n^{\mathrm{Pb}}}{0.2 \, \mathrm{fm}} \right),
      \label{eq:fit_m_DR_Pb} \\
    \frac{M}{M_{\odot}}
    & =
      a_{0, {\mathrm{Sn}}}^M
      +
      a_{1, {\mathrm{Sn}}}^M
      \left( \frac{\Delta R_n^{\mathrm{Sn}}}{0.2 \, \mathrm{fm}} \right),
      \label{eq:fit_m_DR_Sn}
  \end{align}
\end{subequations}
where the adjusting parameters of $a_{i, {\mathrm{Pb}}}^M$ and $a_{i, {\mathrm{Sn}}}^M$
for $ i = 0 $ and $ 1 $ depend on the central density of the corresponding neutron star model,
$ u_{\mathrm{c}} $.
In Fig.~\ref{fig:a_dRM},
we show the dependence of these parameters on $ u_{\mathrm{c}} $,
where the solid lines correspond to the fitting lines given by
\begin{subequations}
  \label{eq:M-DR-a} 
  \begin{align}
    a_{i, {\mathrm{Pb}}}^M
    & =
      \sum_{j = 0}^4
      a_{ij, {\mathrm{Pb}}}^M u_{\mathrm{c}}^j,
      \label{eq:M-DR-Pb-a} \\
    a_{i, {\mathrm{Sn}}}^M
    & =
      \sum_{j = 0}^4
      a_{ij, {\mathrm{Sn}}}^M u_{\mathrm{c}}^j.
      \label{eq:M-DR-Sn-a}
  \end{align}
\end{subequations}
The concrete values of $a_{ij, {\mathrm{Pb}}}^M$ and $a_{ij, {\mathrm{Sn}}}^M$ are listed in Table \ref{tab:aij}.
%
%
\begin{table*}[tb]
  \centering
  \caption{
    The coefficients in Eqs.~(\ref{eq:M-DR-Pb-a}), (\ref{eq:M-DR-Sn-a}),
    (\ref{eq:z-DR-Pb-a}), and (\ref{eq:z-DR-Sn-a}).} 
  \label{tab:aij}
  \begin{ruledtabular}
    \begin{tabular}{lddddd}
      $ j $ & \multicolumn{1}{c}{$ 0 $} & \multicolumn{1}{c}{$ 1 $} & \multicolumn{1}{c}{$ 2 $} & \multicolumn{1}{c}{$ 3 $} & \multicolumn{1}{c}{$ 4 $} \\
      \hline
      $a_{0j, {\mathrm{Pb}}}^M$  & 0.8777    & -2.0127 &  0.9596 & -0.13770  &  0.0047513 \\ 
      $a_{1j, {\mathrm{Pb}}}^M$  & -0.7924   &  1.7396 & -0.3281 & -0.04253  &  0.0117856 \\
      \hline 
      $a_{0j, {\mathrm{Sn}}}^M$  & 1.0850    & -2.4709 &  1.0488 & -0.13010  &  0.0023109 \\ 
      $a_{1j, {\mathrm{Sn}}}^M$  & -0.7604   &  1.6719 & -0.3168 & -0.03827  &  0.0108140 \\ 
      \hline 
      $a_{0j, {\mathrm{Pb}}}^z$  &  0.076109 & -0.2565 &  0.1586 & -0.035961 &  0.0034092 \\ 
      $a_{1j, {\mathrm{Pb}}}^z$  & -0.084495 &  0.2528 & -0.1063 &  0.025645 & -0.0027859 \\ 
      \hline 
      $a_{0j, {\mathrm{Sn}}}^z$  & 0.099032  & -0.3246 &  0.1884 & -0.043693 &  0.0042857 \\ 
      $a_{1j, {\mathrm{Sn}}}^z$  & -0.081280 &  0.2429 & -0.1025 &  0.024997 & -0.0027348 \\ 
    \end{tabular}
  \end{ruledtabular}
\end{table*}
\par
\begin{figure*}[bp]
  \centering
  \includegraphics[width=1.0\linewidth]{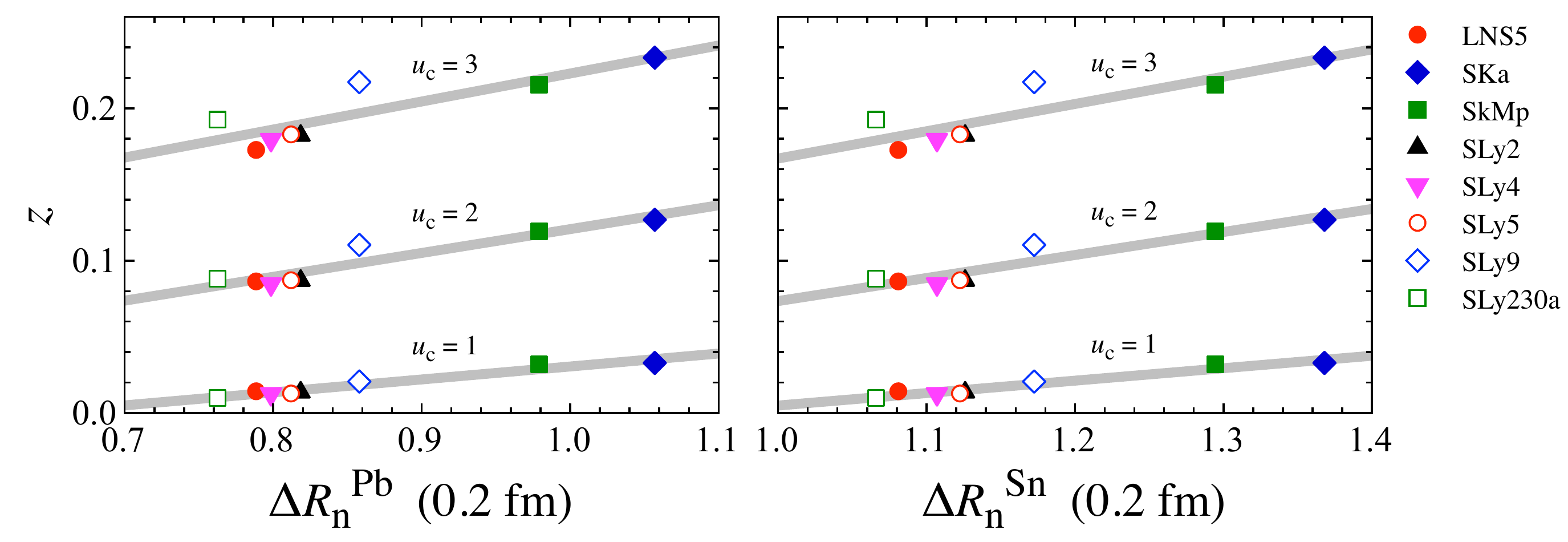}
  \caption{
    The gravitational redshift of the neutron stars constructed with each EOS model
    is shown as a function of $ \Delta R_n^{\mathrm{Pb}} $
    in the left panel and
    $ \Delta R_n^{\mathrm{Sn}}$
    in the right panel for the stellar models with
    $ u_{\mathrm{c}} = 1 $, $ 2 $, and $ 3 $.
    The fitting lines are given by Eqs.~(\ref{eq:fit_z_DR_Pb}) and (\ref{eq:fit_z_DR_Sn}).}
  \label{fig:z-DR}
\end{figure*}

In a similar way, we find that the gravitational redshift, $z$, for the neutron star model with the fixed central density can be expressed as a function of $\Delta R_n$, weakly depending on the EOS models as
\begin{subequations}
  \label{eq:fit_z_DR}
  \begin{align}
    z
    & =
      a_{0, {\mathrm{Pb}}}^z
      +
      a_{1, {\mathrm{Pb}}}^z
      \left( \frac{\Delta R_n^{\mathrm{Pb}}}{0.2 \, \mathrm{fm}} \right),
      \label{eq:fit_z_DR_Pb} \\
    z
    & =
      a_{0, {\mathrm{Sn}}}^z
      +
      a_{1, {\mathrm{Sn}}}^z
      \left( \frac{\Delta R_n^{\mathrm{Sn}}}{0.2\, \mathrm{fm}} \right),
      \label{eq:fit_z_DR_Sn}
  \end{align}
\end{subequations}
where the adjusting parameters of $a_{i, {\mathrm{Pb}}}^z$ and $a_{i, {\mathrm{Sn}}}^z$
for $ i = 0 $ and $ 1 $ depend on the central density of the neutron star.
As an example, we show the results
with $ u_{\mathrm{c}} = 1 $, $ 2 $, and $ 3 $ in Fig.~\ref{fig:z-DR},
where the marks denote the value of $ z $ for the neutron stars constructed with each EOS model,
while the solid lines are the fittings.
Then, as shown in Fig.~\ref{fig:a_dRM},
we can derive the dependence of the adjusting parameters
in Eqs.~(\ref{eq:fit_z_DR_Pb}) and (\ref{eq:fit_z_DR_Sn}) on $ u_{\mathrm{c}} $, as
\begin{subequations}
  \label{eq:z-DR-a}
  \begin{align}
    a_{i, {\mathrm{Pb}}}^z
    & =
      \sum_{j = 0}^4
      a_{ij, {\mathrm{Pb}}}^z u_{\mathrm{c}}^j,
      \label{eq:z-DR-Pb-a} \\
    a_{i, {\mathrm{Sn}}}^z
    & =
      \sum_{j = 0}^4
      a_{ij, {\mathrm{Sn}}}^z u_{\mathrm{c}}^j.
      \label{eq:z-DR-Sn-a}
  \end{align}
\end{subequations}
The solid lines in Fig.~\ref{fig:a_dRM} show the expected values with these fittings.
The concrete values of $a_{ij, {\mathrm{Pb}}}^z$ and $a_{ij, {\mathrm{Sn}}}^z$ are listed in Table \ref{tab:aij}. 
%
\par
\begin{figure*}[tb]
  \centering
  \includegraphics[width=1.0\linewidth]{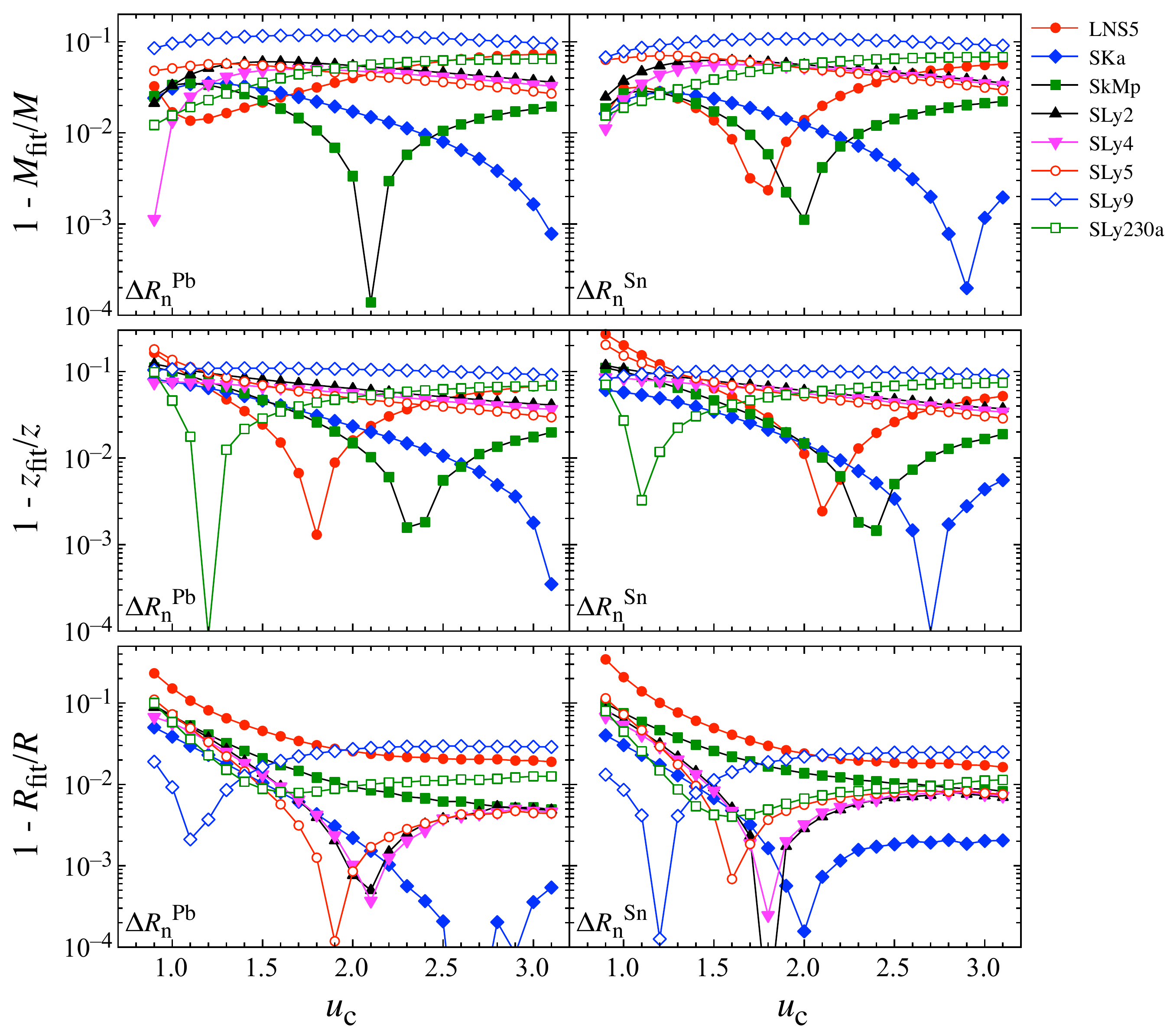}
  \caption{
    Relative deviation of the stellar mass ($M$) and gravitational redshift ($z$)
    estimated with the empirical relations
    (Eqs.~(\ref{eq:fit_m_DR_Pb})--(\ref{eq:z-DR-Sn-a}))
    from those determined as the TOV solutions is shown as a function of $ u_{\mathrm{c}} $.
    The bottom panels correspond to the relative deviation of the stellar radius ($R$) predicted with the empirical formulae for $M$ and $z$ from that determined as the TOV solutions.
    The left and right panels correspond to the results obtained from the empirical formulae as a function of $\Delta R_n^{\mathrm{Pb}}$ and $\Delta R_n^{\mathrm{Sn}}$, respectively.}
  \label{fig:delta-dR}
\end{figure*}

Now, we have derived the empirical formulae expressing $M$ and $z$ as a function of
$ \left( u_{\mathrm{c}}, \Delta R_n^{\mathrm{Pb}} \right) $ or
$ \left( u_{\mathrm{c}}, \Delta R_n^{\mathrm{Sn}} \right) $,
respectively, as Eqs.~(\ref{eq:fit_m_DR}) and (\ref{eq:M-DR-a}) and
Eqs.~(\ref{eq:fit_z_DR}) and (\ref{eq:z-DR-a}).
To see the accuracy of the estimation with these empirical formulae,
we calculate the relative deviation in the mass and its gravitational redshift estimated with the empirical formulae from those as a TOV solution,
and show it in the top and middle panels of Fig.~\ref{fig:delta-dR},
where the left and right panels correspond to the results from the formulae with
$\Delta R_n^{\mathrm{Pb}}$ and $\Delta R_n^{\mathrm{Sn}}$, respectively.
From this figure, one can see that the mass and its gravitational redshift are estimated within $ \lsim 10 \, \% $ accuracy,
using the empirical formula with $\Delta R_n^{\mathrm{Pb}}$ or $\Delta R_n^{\mathrm{Sn}}$.
In addition, in the bottom panel of Fig.~\ref{fig:delta-dR},
we show the relative deviation of the neutron star radius estimated with the empirical formulae for $M$ and $z$ from that determined as a TOV solution.
From this figure, we find that the stellar radius for the neutron star
with $ u_{\mathrm{c}} = 2 $--$ 3 $ can be estimated with the empirical formulae for $M$ and $z$
using $\Delta R_n^{\mathrm{Pb}}$ or $\Delta R_n^{\mathrm{Sn}}$ within a few $\%$ accuracy.
Since these empirical formulae are derived using several EOS models selected in this study, the formulae are applicable only in the range of $\Delta R_n$ given by the adopted EOS models.
That is, 
the empirical formulae are applicable in the range of
$ 0.8 \lsim u_{\mathrm{c}} \lsim 3.0 $
and $ 0.153 \lsim \Delta R_n^{\mathrm{Pb}} \lsim 0.211 \, \mathrm{fm} $
or $0.213 \lsim \Delta R_n^{\mathrm{Sn}} \lsim 0.274 \, \mathrm{fm} $ (e.g., see the horizontal axis in Fig. \ref{fig:M-DR}).

%
\subsection{Empirical relations with $ \alpha_{\mathrm{D}} $}
\label{sec:NSb}
\par
\begin{figure*}[tb]
  \centering
  \includegraphics[width=1.0\linewidth]{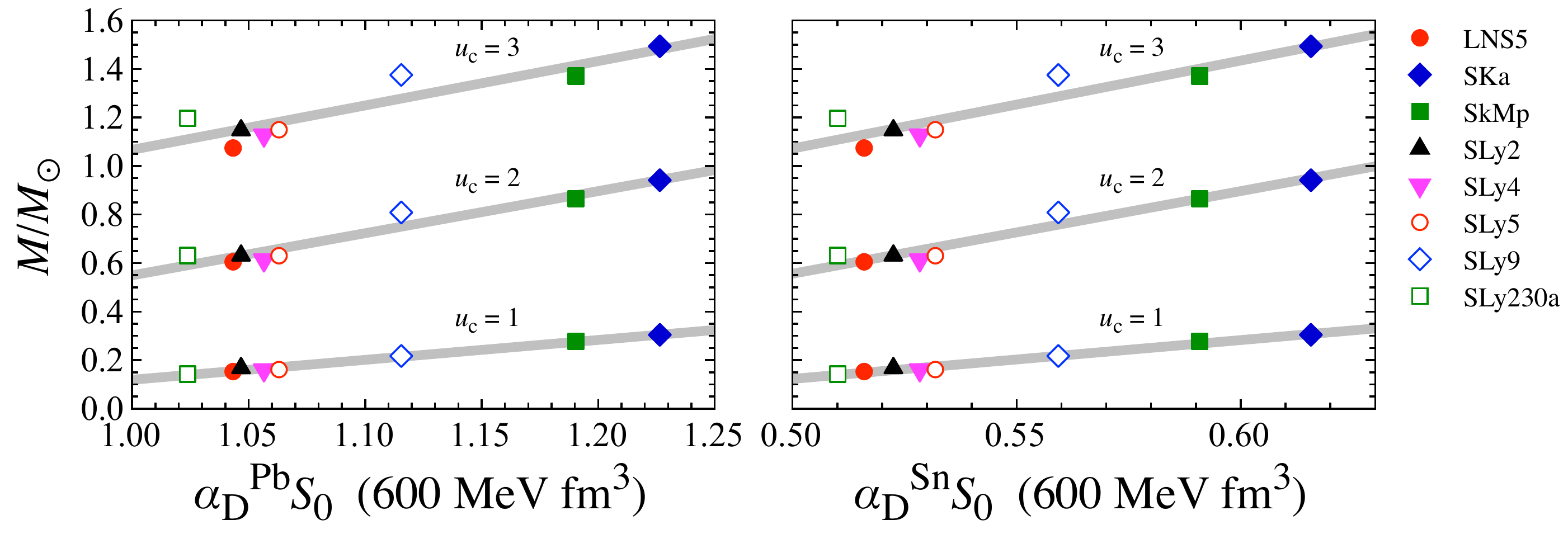}
  \caption{
    The mass of the neutron stars constructed with each EOS model is shown
    as a function of $\alpha_{\mathrm{D}}^{\mathrm{Pb}}S_0$ in the left panel
    and $\alpha_{\mathrm{D}}^{\mathrm{Sn}}S_0$ in the right panel,
    where we show the results for $ u_{\mathrm{c}} = 1 $, $ 2 $, and $ 3 $.
    The fitting lines are given by Eqs.~(\ref{eq:fit_M_DP_Pb}) and (\ref{eq:fit_M_DP_Sn}).}
  \label{fig:M-SDP}
\end{figure*}
\begin{figure}[tb]
  \centering
  \includegraphics[width=1.0\linewidth]{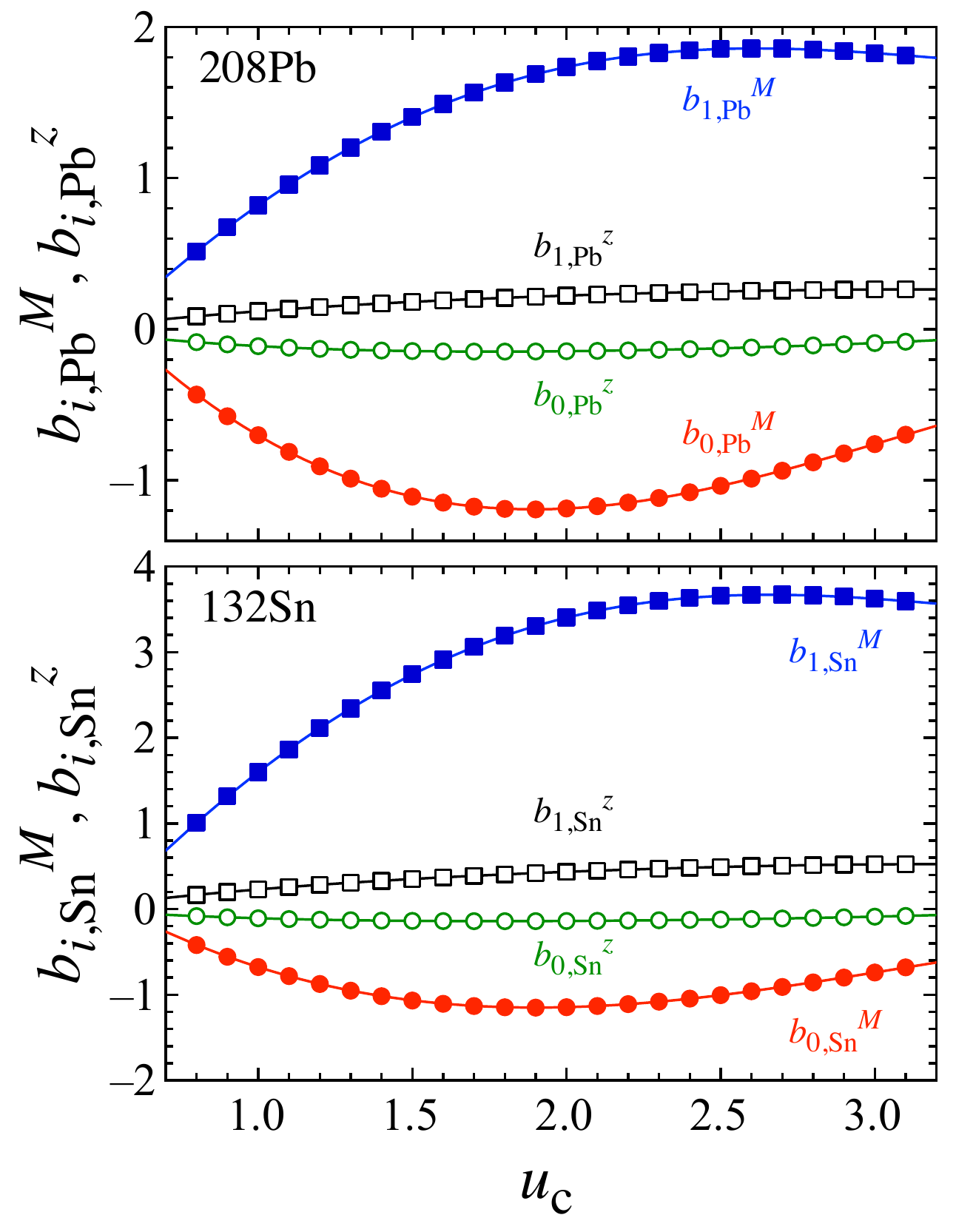}
  \caption{
    The coefficients in Eqs.~(\ref{eq:fit_M_DP_Pb}), (\ref{eq:fit_M_DP_Sn}),
    (\ref{eq:fit_z_DP_Pb}), and (\ref{eq:fit_z_DP_Sn})
    are shown as a function of $u_{\mathrm{c}}$,
    where in the top and bottom panels correspond to the coefficients in the formulae with the data of
    $ {}^{208} \mathrm{Pb} $ and $ {}^{132} \mathrm{Sn} $, respectively,
    while the solid lines denote the fitting lines given by Eqs.~(\ref{eq:M-DP-Pb-b}), (\ref{eq:M-DP-Sn-b}), (\ref{eq:z-DP-Pb-b}), and (\ref{eq:z-DP-Sn-b}).}
  \label{fig:coeffi-M-SDP}
\end{figure}

Next, we consider the derivation of the empirical formulae for $M$ and $z$,
using the dipole polarizability, $ \alpha_{\mathrm{D}} $, for $ {}^{208} \mathrm{Pb} $ and $ {}^{132} \mathrm{Sn} $.
In a similar way to the case with $\Delta R_n$,
we find that the neutron star mass with the fixed central density is strongly correlated to
$ \alpha_{\mathrm{D}} S_0 $, weakly depending on the EOS models.
In Fig.~\ref{fig:M-SDP}, we show the neutron star mass with $ u_{\mathrm{c}} = 1 $, $ 2 $, and $ 3 $
as a function of
$\alpha_{\mathrm{D}}^{\mathrm{Pb}} S_0$ in the left panel and
$\alpha_{\mathrm{D}}^{\mathrm{Sn}} S_0$ in the right panel,
where the solid lines denote the fitting given by
\begin{subequations}
  \label{eq:fit_M_DP}
  \begin{align}
    \frac{M}{M_\odot}
    & =
      b_{0, {\mathrm{Pb}}}^M
      +
      b_{1, {\mathrm{Pb}}}^M
      \left(
      \frac{S_0}{30 \, \mathrm{MeV}}
      \frac{\alpha_{\mathrm{D}}^{\mathrm{Pb}}}{20 \, \mathrm{fm}^3}
      \right),
      \label{eq:fit_M_DP_Pb} \\
    \frac{M}{M_\odot}
    & =
      b_{0, {\mathrm{Sn}}}^M
      +
      b_{1, {\mathrm{Sn}}}^M
      \left(
      \frac{S_0}{30 \, \mathrm{MeV}}
      \frac{\alpha_{\mathrm{D}}^{\mathrm{Sn}}}{20 \, \mathrm{fm}^3}
      \right).
      \label{eq:fit_M_DP_Sn}
  \end{align}
\end{subequations}
The coefficients in these fittings depend on the value of $u_{\mathrm{c}}$,
and we can derive their fitting as
\begin{subequations}
  \label{eq:M-DP-b}
  \begin{align}
    b_{i, {\mathrm{Pb}}}^M
    & =
      \sum_{j = 0}^4
      b_{ij, {\mathrm{Pb}}}^M u_{\mathrm{c}}^j,
      \label{eq:M-DP-Pb-b} \\
    b_{i, {\mathrm{Sn}}}^M
    & =
      \sum_{j = 0}^4
      b_{ij, {\mathrm{Sn}}}^M u_{\mathrm{c}}^j,
      \label{eq:M-DP-Sn-b} 
  \end{align}
\end{subequations}
with which the expected values are shown in Fig.~\ref{fig:coeffi-M-SDP} with the solid lines.
The concrete values of $b_{ij, {\mathrm{Pb}}}^M$ and $b_{ij, {\mathrm{Sn}}}^M$ for
$i = 0$, $ 1 $ and $ j = 0 $--$ 4 $ are listed in Table \ref{tab:bij}.
%
%
\begin{table*}[tb]
  \centering
  \caption{
    The coefficients in Eqs.~(\ref{eq:M-DP-Pb-b}), (\ref{eq:M-DP-Sn-b}), (\ref{eq:z-DP-Pb-b}), and (\ref{eq:z-DP-Sn-b}).} 
  \label{tab:bij}
  \begin{ruledtabular}
    \begin{tabular}{cddddd}
      $ j $ & \multicolumn{1}{c}{0} & \multicolumn{1}{c}{1} & \multicolumn{1}{c}{2} & \multicolumn{1}{c}{3} & \multicolumn{1}{c}{4} \\
      \hline
      $b_{0j, {\mathrm{Pb}}}^M$  &   1.4048 & -3.1188 &  1.0989 & -0.08108  & -0.006288  \\ 
      $b_{1j, {\mathrm{Pb}}}^M$  &  -1.1036 &  2.3765 & -0.3867 & -0.08429  &  0.019227  \\ 
      \hline 
      $b_{0j, {\mathrm{Sn}}}^M$  &   1.3332 & -2.9720 &  1.0432 & -0.07696  & -0.0057868 \\
      $b_{1j, {\mathrm{Sn}}}^M$  &  -2.0784 &  4.4887 & -0.6693 & -0.17770  &  0.037747  \\ 
      \hline 
      $b_{0j, {\mathrm{Pb}}}^z$  &   0.1340 & -0.4237 &  0.2239 & -0.051039 &  0.0050455 \\ 
      $b_{1j, {\mathrm{Pb}}}^z$  &  -0.1191 &  0.3508 & -0.1430 &  0.033876 & -0.0036785 \\ 
      \hline 
      $b_{0j, {\mathrm{Sn}}}^z$  &  0.1279  & -0.4069 &  0.2173 & -0.050428 &  0.0050662 \\ 
      $b_{1j, {\mathrm{Sn}}}^z$  & -0.2264  &  0.6699 & -0.2725 &  0.066119 & -0.0073334 \\ 
    \end{tabular}
  \end{ruledtabular}
\end{table*}
\par
\begin{figure*}[tb]
  \centering
  \includegraphics[width=1.0\linewidth]{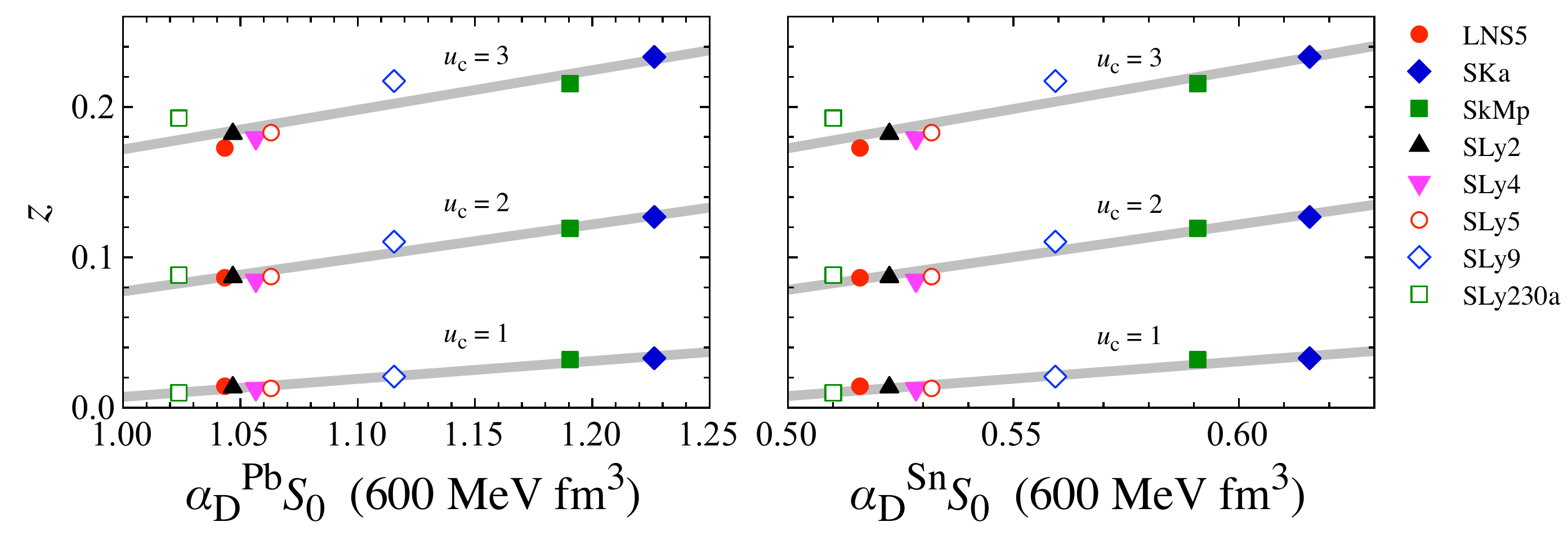}
  \caption{
    The gravitational redshift of the neutron stars constructed with each EOS model
    is shown as a function of $ \alpha_{\mathrm{D}}^{\mathrm{Pb}} S_0 $ in the left panel
    and $ \alpha_{\mathrm{D}}^{\mathrm{Sn}} S_0 $ in the right panel,
    where we show the results for $ u_{\mathrm{c}} = 1 $, $ 2 $, and $ 3 $.
    The fitting lines are given by Eqs.~(\ref{eq:fit_z_DP_Pb}) and (\ref{eq:fit_z_DP_Sn}).}
  \label{fig:z-SDP}
\end{figure*}

Moreover, we also find that the gravitational redshift of the neutron star with the fixed central density is strongly associated with $\alpha_{\mathrm{D}} S_0$, weakly depending on the EOS models.
In fact, as shown in Fig.~\ref{fig:z-SDP},
it can be expressed as a function of $\alpha_{\mathrm{D}} S_0$ for $ {}^{208} \mathrm{Pb} $ and $ {}^{132} \mathrm{Sn} $, such as
\begin{subequations}
  \label{eq:fit_z_DP}
  \begin{align}
    z
    & =
      b_{0, {\mathrm{Pb}}}^z
      +
      b_{1, {\mathrm{Pb}}}^z
      \left(
      \frac{S_0}{30 \, \mathrm{MeV}}
      \frac{\alpha_{\mathrm{D}}^{\mathrm{Pb}}}{20 \, \mathrm{fm}^3}
      \right),
      \label{eq:fit_z_DP_Pb} \\
    z
    & =
      b_{0, {\mathrm{Sn}}}^z
      +
      b_{1, {\mathrm{Sn}}}^z
      \left(
      \frac{S_0}{30 \, \mathrm{MeV}}
      \frac{\alpha_{\mathrm{D}}^{\mathrm{Sn}}}{20 \, \mathrm{fm}^3}
      \right),
      \label{eq:fit_z_DP_Sn}
  \end{align}
\end{subequations}
where the coefficients of $b_{i, {\mathrm{Pb}}}^z$ and $b_{i, {\mathrm{Sn}}}^z$ should be expressed with $u_{\mathrm{c}}$.
We show such dependence in Fig.~\ref{fig:coeffi-M-SDP},
where the solid lines are the fittings given by
\begin{subequations}
  \label{eq:z-DP-b}
  \begin{align}
    b_{i, {\mathrm{Pb}}}^z
    & =
      \sum_{j = 0}^4
      b_{ij, {\mathrm{Pb}}}^z u_{\mathrm{c}}^j,
      \label{eq:z-DP-Pb-b} \\
    b_{i, {\mathrm{Sn}}}^z
    & =
      \sum_{j = 0}^4
      b_{ij, {\mathrm{Sn}}}^z u_{\mathrm{c}}^j.
      \label{eq:z-DP-Sn-b}
  \end{align}
\end{subequations}
The concrete values of $b_{ij, {\mathrm{Pb}}}^z$ and $b_{ij, {\mathrm{Sn}}}^z$
for $ i = 0 $, $ 1 $ and $ j = 0 $--$ 4 $ are listed in Table \ref{tab:bij}. 
%
\par
\begin{figure*}[tb]
  \centering
  \includegraphics[width=1.0\linewidth]{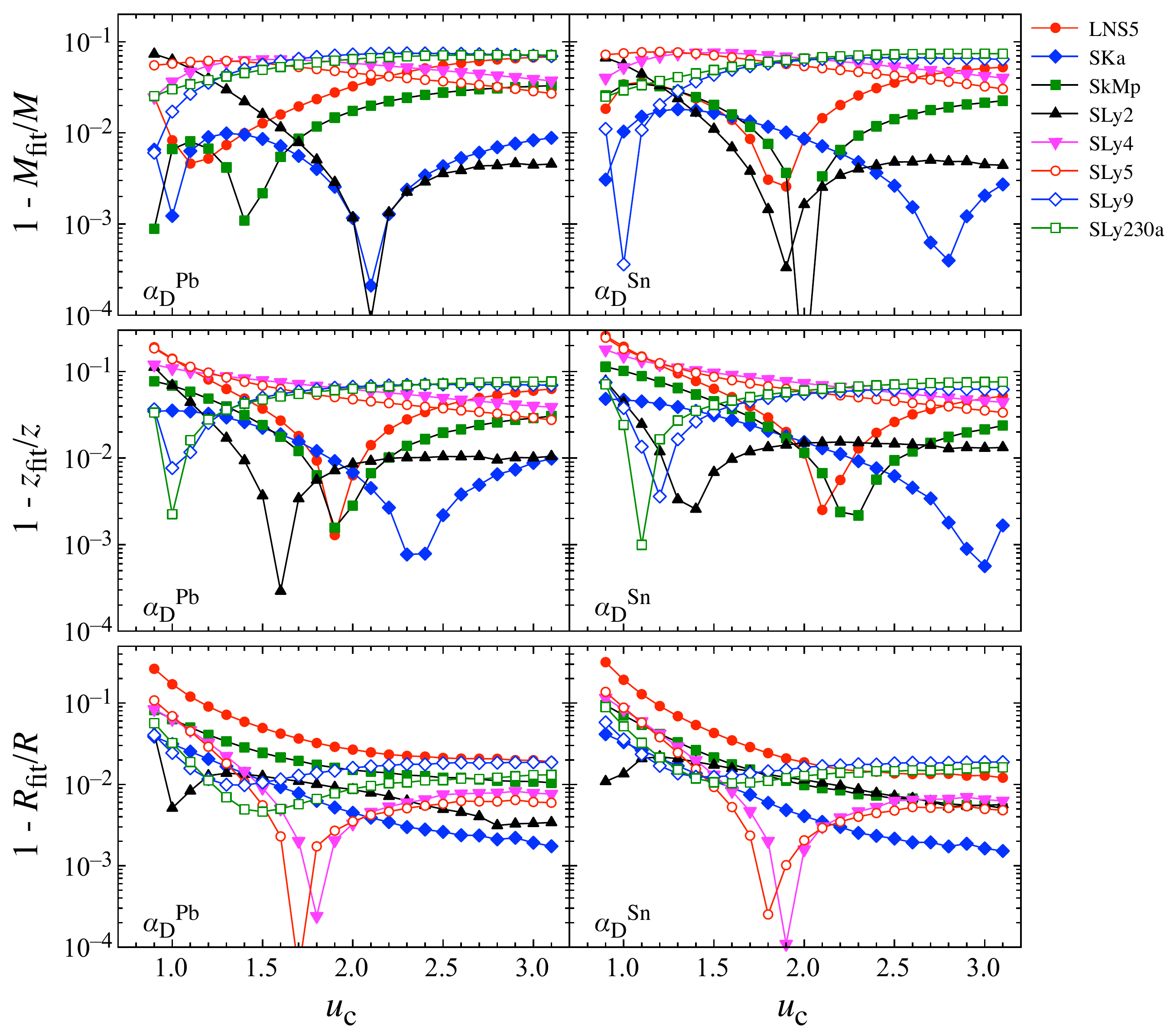}
  \caption{
    Same as in Fig.~\ref{fig:delta-dR},
    but with the empirical formulae using $\alpha_{\mathrm{D}}^{\mathrm{Pb}}$ in the left panel and $\alpha_{\mathrm{D}}^{\mathrm{Sn}}$ in the right panel.}
  \label{fig:delta-DP}
\end{figure*}

Now, we have newly obtained the empirical formulae expressing $M$ and $z$ as a function of
$ \left( u_{\mathrm{c}}, \alpha_{\mathrm{D}}^{\mathrm{Pb}}S_0 \right) $ or
$ \left( u_{\mathrm{c}}, \alpha_{\mathrm{D}}^{\mathrm{Sn}}S_0 \right) $,
respectively, as Eqs.~(\ref{eq:fit_M_DP}) and (\ref{eq:M-DP-b}) and
Eqs.~(\ref{eq:fit_z_DP}) and (\ref{eq:z-DP-b}).
In the top and middle panels of Fig.~\ref{fig:delta-DP},
we show the relative deviation of the estimation of mass and gravitational redshift with empirical formulae from those determined by integrating the TOV equations.
From this figure,
we find that the mass and gravitational redshift can be estimated within $\approx 8 \, \% $ accuracy from the empirical formulae with $ \alpha_{\mathrm{D}}^{\mathrm{Pb}}S_0 $ or $ \alpha_{\mathrm{D}}^{\mathrm{Sn}}S_0 $.
Additionally, one can estimate the stellar radius by combining these formulae for $M$ and $z$,
whose relative deviation from the stellar radius as the TOV solution is shown in the bottom panel of Fig.~\ref{fig:delta-DP}.
From this figure, we find the stellar radius can be estimated within $\approx 2 \, \% $ error for the neutron star for $ u_{\mathrm{c}} = 2 $--$ 3 $.
These empirical formulae are applicable in the range of
$ 0.8 \lsim u_{\mathrm{c}} \lsim 3.0 $ and
$ 1.02 \lsim \alpha_{\mathrm{D}}^{\mathrm{Pb}} S_0 / \left( 600 \, \mathrm{MeV} \, \mathrm{fm}^3 \right) \lsim 1.23 $
or
$ 0.51 \lsim \alpha_{\mathrm{D}}^{\mathrm{Sn}} S_0 / \left( 600 \, \mathrm{MeV} \, \mathrm{fm}^3 \right) \lsim 0.62$.
%
%
\section{Neutron star mass and radius relation}
\label{sec:MR}
\par
\begin{figure*}[tb]
  \centering
  \includegraphics[width=1.0\linewidth]{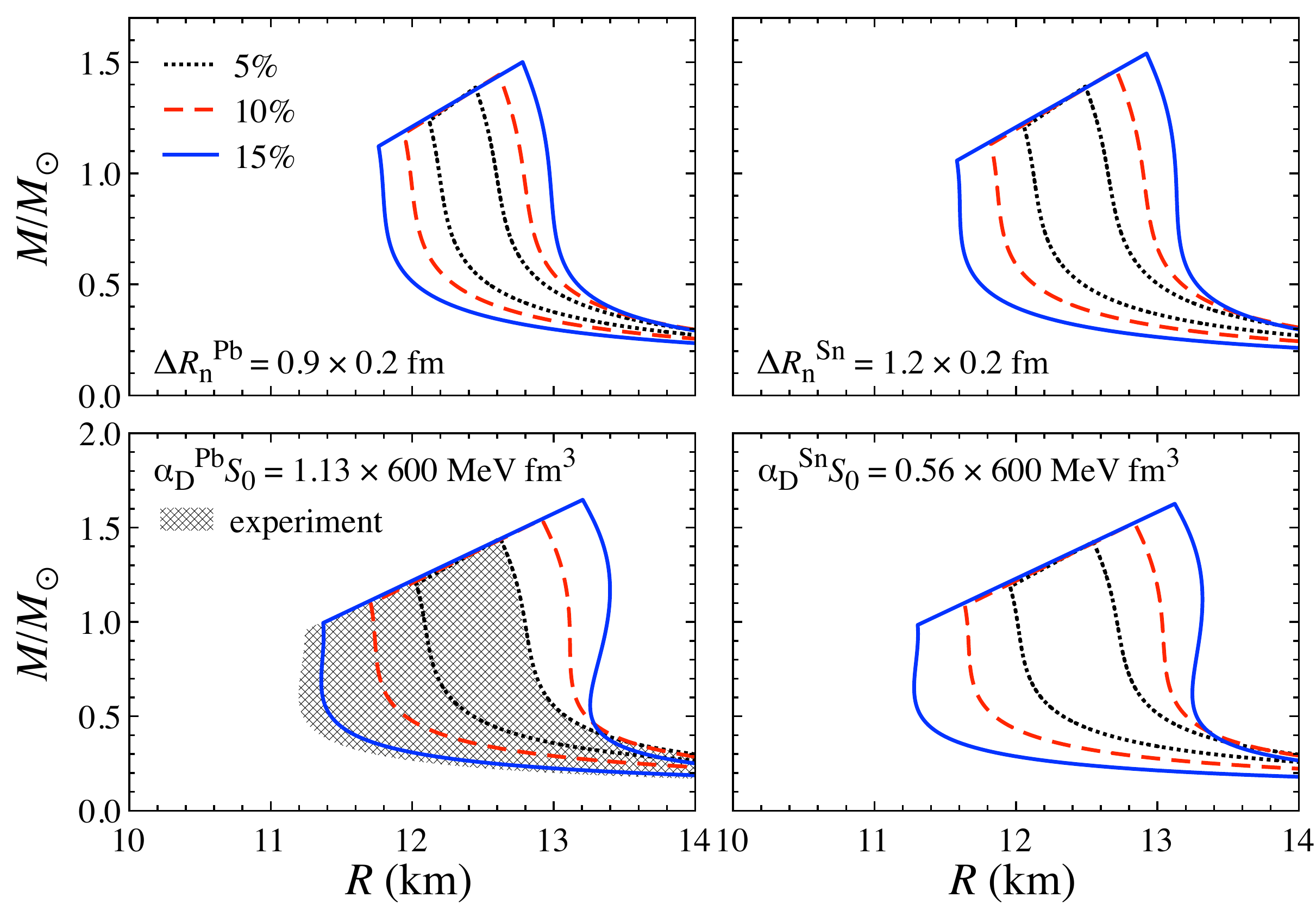}
  \caption{
    The neutron star mass and radius predicted with the empirical formulae with $\Delta R_n^{\mathrm{Pb}}$ (top-left),
    $\Delta R_n^{\mathrm{Sn}}$ (top-right),
    $\alpha_{\mathrm{D}}^{\mathrm{Pb}}S_0$ (bottom-left),
    and $\alpha_{\mathrm{D}}^{\mathrm{Sn}}S_0$ (top-left),
    where we assume that
    $ \Delta R_n^{\mathrm{Pb}}/ \left( 0.2 \, \mathrm{fm} \right) = 0.9 $,
    $ \Delta R_n^{\mathrm{Sn}}/ \left( 0.2 \, \mathrm{fm} \right) = 1.2 $,
    $ \alpha_{\mathrm{D}}^{\mathrm{Pb}}S_0/ \left( 600 \, \mathrm{MeV} \, \mathrm{fm}^3 \right) = 1.13 $,
    $ \alpha_{\mathrm{D}}^{\mathrm{Sn}}S_0/ \left( 600 \, \mathrm{MeV} \, \mathrm{fm}^3 \right) = 0.56 $
    as their test values.
    The dotted, dashed, and solid lines denote the region of the neutron star mass and radius with
    $ \pm 5 \, \% $, $ \pm 10 \, \% $, and $ \pm 15 \, \% $ deviation from the test values. 
    In the left-bottom panel, we also show the shaded region as the predicted region from the experimental value of $\alpha_{\text{D}}^{\text{Pb}} = 20.1 (6) \, \mathrm{fm}^3 $ together with the fiducial value of $S_0$ (see text for details).}
  \label{fig:MR}
\end{figure*}

Using the empirical formula derived in this study,
we see how the neutron star mass and radius depend on the experimental observables,
i.e., $\Delta R_n$ and $\alpha_{\mathrm{D}} S_0$.
For this purpose, we assume that
$ \Delta R_n^{\mathrm{Pb}} / \left( 0.2 \, \mathrm{fm} \right) = 0.9 $,
$ \Delta R_n^{\mathrm{Sn}} / \left( 0.2 \, \mathrm{fm} \right) = 1.2 $,
$ \alpha_{\mathrm{D}}^{\mathrm{Pb}} S_0 / \left( 600 \, \mathrm{MeV} \, \mathrm{fm}^3 \right) = 1.13$,
and 
$ \alpha_{\mathrm{D}}^{\mathrm{Sn}} S_0 / \left( 600 \, \mathrm{MeV} \, \mathrm{fm}^3 \right) = 0.56 $
as their test values here.
These values are around the central values in the range of corresponding variables with the EOS models adopted in this study
(see the horizontal axis in Figs.~\ref{fig:M-DR}, \ref{fig:z-DR}, \ref{fig:M-SDP}, and \ref{fig:z-SDP}).
Then, in Fig.~\ref{fig:MR}, we show the neutron star mass and radius predicted from the empirical relations,
adopting the error of $ \pm 5 \, \% $, $ \pm 10 \, \% $, and $ \pm 15 \, \% $ from the test values.
  In addition, the experimental value of $\Delta R_n^{\text{Pb}}$ 
  is known via PREX-II, i.e., $\Delta R_n^{\text{Pb}} = 0.283 \pm 0.071 \, \mathrm{fm} $~\cite{PREXII},
  but this constraint is out of the range in which our empirical relation is applicable.
  On the other hand, the experimental value of
  $ \alpha_{\text{D}}^{\text{Pb}} $ is $\alpha_{\text{D}}^{\text{Pb}} = 20.1(6) \, \mathrm{fm}^3 $~\cite{10.1143/PTPS.196.166}, 
  which leads $ 0.94 \lsim \alpha_{\text{D}}^{\text{Pb}} S_0 / \left( 600 \, \mathrm{MeV} \, \mathrm{fm}^3 \right) \lsim 1.18 $,
  assuming that $ S_0 \approx 31.6 \pm 2.7 \, \mathrm{MeV} $~\cite{BALi19}.
  Since this is more or less inside the applicable range, we also show the predicted region in the neutron star mass and radius, using this experimental value, in the left-bottom panel in Fig.~\ref{fig:MR}.
From this figure, one can observe that the predicted neutron star mass and radius strongly depend on the experimental observables,
even if one assumes the same errors in $ \Delta R_n $ and $ \alpha_{\mathrm{D}} S_0$.
In fact, it seems that one can well predict the neutron star mass and radius,
using the data of $\Delta R_n$ for $ {}^{208} \mathrm{Pb} $.
For example, once one would determine the value of $\Delta R_n$ for $ {}^{208} \mathrm{Pb} $ within $10 \, \%$ error, neutron star radius may be determined a few $\%$ accuracy. 
\par
To understand this situation,
we see the EOS dependence of $\Delta R_n^{\mathrm{Pb}}$, $\Delta R_n^{\mathrm{Sn}}$,
$\alpha_{\mathrm{D}}^{\mathrm{Pb}}S_0$, and $\alpha_{\mathrm{D}}^{\mathrm{Sn}}S_0$.
From Figs.~\ref{fig:M-DR}, \ref{fig:z-DR}, \ref{fig:M-SDP}, and \ref{fig:z-SDP},
one can observe the minimum and maximum values of
$\Delta R_n^{\mathrm{Pb}}$, $\Delta R_n^{\mathrm{Sn}}$, $\alpha_{\mathrm{D}}^{\mathrm{Pb}}S_0$, and $\alpha_{\mathrm{D}}^{\mathrm{Sn}}S_0$ are given by SLy230a and SKa, respectively.
Now, to see how these variables strongly depend on the EOS models, we calculate their relative range through
\begin{equation}
  \label{eq:deviation}
  \delta B
  =
  \frac{2 \left( B_{\mathrm{SKa}} - B_{\mathrm{SLy230a}} \right)}{B_{\mathrm{SKa}} + B_{\mathrm{SLy230a}}},
\end{equation}
where $B$ denotes the variables of
$\Delta R_n^{\mathrm{Pb}}$, $\Delta R_n^{\mathrm{Sn}}$,
$\alpha_{\mathrm{D}}^{\mathrm{Pb}}S_0$, and $\alpha_{\mathrm{D}}^{\mathrm{Sn}}S_0$.
One can get
$ \delta \left( \Delta R_n^{\mathrm{Pb}} \right) = 0.324 $,
$ \delta \left( \Delta R_n^{\mathrm{Sn}} \right) = 0.248 $,
$ \delta \left( \alpha_{\mathrm{D}}^{\mathrm{Pb}}S_0 \right) = 0.180 $, and
$ \delta \left( \alpha_{\mathrm{D}}^{\mathrm{Sn}}S_0 \right) = 0.186 $.
This means that the EOS dependence of $\Delta R_n^{\mathrm{Pb}}$ is stronger than those of
$\alpha_{\mathrm{D}}^{\mathrm{Pb}}S_0$ or $\alpha_{\mathrm{D}}^{\mathrm{Sn}}S_0$.
That is, if one considers the same errors in the experimental observables,
such as $ \pm 15 \, \% $, $ \alpha_{\mathrm{D}}^{\mathrm{Pb}}S_0 $ or $ \alpha_{\mathrm{D}}^{\mathrm{Sn}}S_0 $
easily gets out from the applicable range.
Actually, the values of $\alpha_{\mathrm{D}}^{\mathrm{Pb}}S_0$ and $\alpha_{\mathrm{D}}^{\mathrm{Sn}}S_0$
with $ \pm 15 \, \% $ errors from the test values are out of the applicable range.
This is a reason why the neutron star mass and radius are relatively well predicted with the empirical formula with $\Delta R_n^{\mathrm{Pb}}$.
Conversely, we may conclude that it is difficult to constrain the neutron star mass and radius from the dipole polarizability.
\par
This tendency can be understood as follows:
The neutron-skin thickness is basically determined by the isovector properties of the effective interaction, and accordingly,
the neutron-skin thickness strongly correlates with the neutron excess $ \left( N - Z \right) / A $ (equivalent to an asymmetric parameter $\alpha$), where $N$, $Z$, and $A$ denote the neutron number, proton number, and atomic mass number of each nucleus.
Since the neutron excess of $ {}^{208} \mathrm{Pb} $ is smaller than $ {}^{132} \mathrm{Sn} $,
the absolute value of $ \Delta R_n $ for $ {}^{208} \mathrm{Pb} $ is smaller than that for $ {}^{132} \mathrm{Sn} $. On the other hand,
the deviation of $ \Delta R_n $ among the models in $ {}^{208} \mathrm{Pb} $, i.e., $ y $-axis of Fig.~\ref{fig:DRn-L}, is, eventually, almost the same as those in $ {}^{132} \mathrm{Sn} $.
Accordingly, $ \delta \left( \Delta R_n^{\mathrm{Pb}} \right) $ becomes larger than $ \delta \left(\Delta R_n^{\mathrm{Sn}} \right) $,
although the accuracy of $ \Delta R_n^{\mathrm{Pb}} $ is expected to be better than $ \Delta R_n^{\mathrm{Sn}} $
since $ {}^{132} \mathrm{Sn} $ is an exotic nucleus.
\par
In contrast, $ \alpha_{\mathrm{D}} $ is associated with the isoscalar properties of the effective interaction;
indeed, $ \alpha_{\mathrm{D}} S_0$ correlates with $ A \left\langle r^2 \right\rangle $
with the mean-square radius of the nucleus $ \left\langle r^2 \right\rangle $~\cite{
  Myers1974Ann.Phys.84_186,
  Meyer1982Nucl.Phys.A385_269,
  RocaMaza13}.
Eventually, the deviation of $ \alpha_{\mathrm{D}} S_0 $ among the models, i.e., $ y $-axis of Fig.~\ref{fig:alphaS-L}, scales with the absolute value.
Isoscalar properties of the effective interaction are determined better than isovector ones.
Thus, $ \delta \left( \alpha_{\mathrm{D}} S_0 \right) $ is smaller than $ \delta \left( \Delta R_n \right) $.
%
\par
\begin{figure}[tb]
  \centering
  \includegraphics[width=1.0\linewidth]{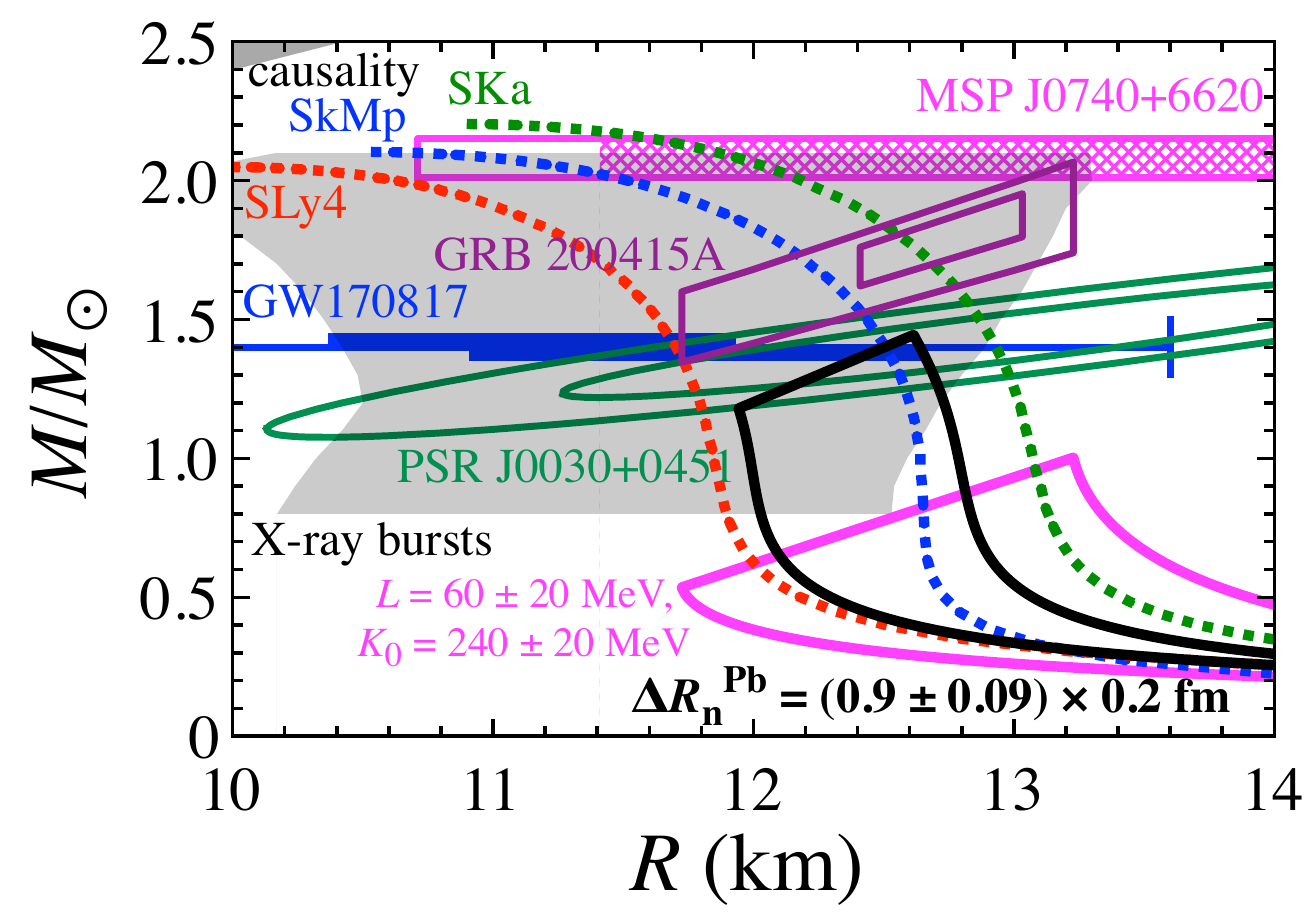}
  \caption{
    The neutron star mass and radius predicted with our empirical formula given by
    Eqs.~(\ref{eq:fit_m_DR_Pb}), (\ref{eq:M-DR-Pb-a}),
    (\ref{eq:fit_z_DR_Pb}), and (\ref{eq:z-DR-Pb-a}),
    assuming that $ \Delta R_n^{\mathrm{Pb}} = \left( 0.9 \pm 0.09 \right) \times 0.2 \, \mathrm{fm} $.
    In the figure, we also plot the constraints from the astronomical observations in GW170817, by NICER (PSR J0030+451 and MSP J0740+6620), via X-ray bursts, and with the magnetar QPOs (GRB 200415A), together with the mass and radius region predicted with the mass formulae with the nuclear saturation parameters, assuming $ L = 60 \pm 20 $ and $ K_0 = 240 \pm 20 \, \mathrm{MeV} $.
    The top-left region is excluded from the causality. For reference, the mass and radius for the neutron stars constructed with the SKa, SkMp, and SLy4 EOSs are shown with the dotted lines. See the text for the details.}
  \label{fig:MR1}
\end{figure}

Finally, in Fig.~\ref{fig:MR1}, we compare the expected region of neutron star mass and radius with the resultant empirical formula of $\Delta R_n^{\mathrm{Pb}}$, i.e.,
Eqs.~(\ref{eq:fit_m_DR_Pb}), (\ref{eq:M-DR-Pb-a}), (\ref{eq:fit_z_DR_Pb}), and (\ref{eq:z-DR-Pb-a}),
assuming that
$ \Delta R_n^{\mathrm{Pb}} = \left( 0.9 \pm 0.09 \right) \times 0.2 \, \mathrm{fm} $
($ \pm 10 \, \% $ deviation)
to the other constraints.
As shown in Fig.~\ref{fig:MR},
for the neutron star model predicted with the empirical formula of $\Delta R_n^{\mathrm{Pb}}$,
we only plot the stellar model whose central density is up to threefold the saturation density.
In the same figure, we show the constraints on the neutron star mass and radius obtained from the various astronomical observations; the gravitational wave observations in the GW170817 event,
i.e., the $1.4M_\odot$ neutron star radius is less than $ 13.6 \, \mathrm{km} $ \cite{Annala18},
whose constraint may become more severe by combining with the multimessenger observations and nuclear theory \cite{Capano20,Dietrich20};
the x-ray observations via NICER for PSR J0030+0451 \cite{Riley19,Miller19}
and MSP J0740+6620 \cite{Riley21,Miller21};
the observations of x-ray burst through the theoretical models \cite{Steiner13};
and the identification of the magnetar quasi-periodic oscillations observed in GRB 200415A with the crustal torsional oscillations \cite{SK23}.
As the theoretical constraint, the top-left region can be excluded from the causality \cite{Lattimer12}.
We also show the mass and radius region predicted with the empirical formulae with the nuclear saturation parameters,
i.e., $ \eta = \left( K_0 L^2 \right)^{1/3} $ \cite{SIOO14},
assuming that $ L = 60 \pm 20 $ and $ K_0 = 240 \pm 20 \, \mathrm{MeV} $ as their fiducial values,
where the central density is considered to be less than twice the saturation density.
Furthermore, for reference, the neutron star models constructed with some of the EOS models adopted in this study, such as the SKa, SkMp, and SLy4, are also shown with dotted lines. 
%
%
\section{Conclusion}
\label{sec:Conclusion}
\par
The nuclear saturation parameters must be important parameters characterizing the EOS models.
But, they are usually constrained from the experimental data through a kind of theoretical model.
To avoid such a circumvention way and to directly discuss the neutron star properties,
such as the mass and radius,
with the experimental data, we derive the empirical formulae expressing the neutron star mass and its gravitational redshift,
as a function of the normalized central density,
$u_{\mathrm{c}}=\rho_{\mathrm{c}}/\rho_0$,
and neutron-skin thickness or the dipole polarizability for $ {}^{208} \mathrm{Pb} $ or $ {}^{132} \mathrm{Sn} $.
These formulae can predict the neutron star mass and its gravitational redshift within $ \approx 10 \, \% $ accuracy,
while the stellar radius is estimated within a few $\%$ accuracy by combining the resultant empirical formulae.
Then, using the empirical formulae, we see how the neutron star mass and radius depend on the experimental data,
i.e., the neutron-skin thickness and dipole polarizability of $ {}^{208} \mathrm{Pb} $ or $ {}^{132} \mathrm{Sn} $.
As a result, we find that the neutron star mass and radius are relatively more sensitive to the data of the neutron-skin thickness $ {}^{208} \mathrm{Pb} $,
while they seem to be less sensitive to the dipole polarizability.
As an example, we show that the neutron star radius could be determined within a few $\%$ accuracy once the neutron-skin thickness $ {}^{208} \mathrm{Pb} $ would be determined within $ 10 \, \% $ errors.
In this study, we successfully derive the empirical formulae expressing the neutron star mass and its gravitational redshift,
but the applicable range may not be so wide.
This is because the sample number of the EOS models is not so much, due to the lack of availability of the EOS models.
To extend the applicable range, we will consider if we could update the empirical formulae by collecting a variety of EOS models as possible. 
%
\acknowledgments
HS is grateful to Susumu Shimoura for his valuable comments. 
TN is grateful to Li-Gang Cao for his comments on the LNS functional.
This work is supported in part by Japan Society for the Promotion of Science (JSPS) KAKENHI Grant Numbers 
JP19KK0354,  
JP21H01088,  
and 
JP22K20372,   
by Pioneering Program of RIKEN for Evolution of Matter in the Universe (r-EMU),
by the RIKEN Special Postdoctoral Researchers Program, 
and by the Science and Technology Hub Collaborative Research Program from RIKEN Cluster for Science, Technology and Innovation Hub (RCSTI). 
The numerical calculations were partly performed on cluster computers at the RIKEN iTHEMS program.
%
%
%
%

\end{document}